# PERFORMANCE BOUNDS OF NANOPARTICLES LADEN VOLUMETRIC ABSORPTION SOLAR THERMAL PLATFORMS IN LAMINAR FLOW REGIME

Apoorva Singh[a], Manish Kumar[b], and Vikrant Khullar[a,*]

[a]Mechanical Engineering Department, Thapar Institute of Engineering and Technology, Patiala-147004, Punjab, India
[b]Department of Mechanical Engineering, Malaviya National Institute of Technology Jaipur, Rajasthan, India
*Corresponding author. Email address: vikrant.khullar@thapar.edu

## Abstract
Recent success in synthesizing thermally stable nanofluids at low costs is a significant breakthrough in the evolution of volumetric absorption based solar thermal systems. However, we have yet not been able to clearly identify the range of operating and design parameters in which volumetric absorption could prove to be beneficial. One of the key reasons being that we have not been able to fully understand the heat transfer mechanisms involved in these novel systems. The present work takes a few steps further in this direction wherein we have developed a comprehensive and mechanistic theoretical framework which is robust enough to account for coupled transport phenomena and orders of magnitudes of operating parameters for host of receiver design configurations. Moreover, we have also modeled equivalent surface absorption based systems to provide a comparison between volumetric and surface absorption processes under similar operating conditions. Performance characteristics reveal that particularly at high solar concentration ratios, volumetric absorption-based receivers could have 35% - 49% higher thermal efficiencies compared to their surface absorption-based counterparts. Finally, the present work serves to define optimal performance domains of these solar thermal systems, particularly in the laminar flow regime ($200 < Re < 1600$) and over a wide range of solar concentration ratios (5-100) and inlet fluid temperatures (293-593K).

## 1. Introduction
In present times, energy security has emerged as one of the major challenges to achieve sustainable development. As per IEA estimates approximately 82% of the total energy demand is met through fossils fuels (i.e. oil, natural gas, and coal) [1]. While, the ever-growing reliance upon fossil fuels to meet the increasing needs of modern world has led to drastic deterioration of the environment. On the other hand, the reserves of fossil fuels are limited. Further, with more and more stringent regulations coming up, use of renewable sources of energy is now considered to be one of the most viable options to not just limit the use of fossil fuels but also to replace them. In this context, solar energy is considered as the most abundant renewable energy resource and possesses the highest technical feasible potential (about 60TW) among all renewable energy resources [2]. However, currently solar energy accounts for only ~0.04 % of total energy demand globally due to lower efficiencies and high costs as compared to fossil-fuel based technologies [1, 3].

In solar energy harvesting, one of the most efficient practices is the use of the concentrated solar thermal technologies. In this, the surface (solar selective/black) absorbs the concentrated solar radiation, and subsequently transfers the absorbed energy to the fluid flowing underneath. While this mechanism is efficient at converting incident radiation to thermal energy; it does not as efficiently transfer this energy to the heat carrier fluid owing to thermal resistance between the surface and the fluid - limiting the transfer rate of energy. Moreover, since the temperature of the surface is the highest; large energy loss is incurred in the form of radiative losses. These limitations could be addressed by using volumetric absorption-based receivers (VARs); herein, sunlight is directly (and volumetrically) absorbed as well as transported by the working fluid itself without the need of any intervening surface. This results in efficient photo-thermal energy conversion of the incident sunlight into the thermal energy of the working fluid [4-8].

Stability of nanoparticle dispersions under real world conditions and optimizing receiver design to achieve higher thermal efficiencies have been the two most active domains of research in the development of nanofluid based volumetric absorption solar thermal platforms [9-15].

Given the broad absorption characteristics of carbon nanostructures laden fluids, these have emerged as potential heat transfer fluids for volumetric absorption solar thermal platforms [16, 17]. Moreover, recent breakthroughs in engineering thermally stable nanofluids at low costs have to a large extent addressed the operational issues in employing nanofluids in real world volumetric absorption solar thermal platforms [18-21]. However, we are still to arrive at an optimum receiver design. This may be attributed to the fact that that we have not been able to fully model the heat transfer mechanisms involved in these thermal systems. These enticingly simple volumetric absorption systems are in reality far more complex and difficult to model as it involves wide range of coupled physics encompassing heat transfer, fluid mechanics, nanotechnology, optics and material science.

Modeling of such receivers as reported in the scientific literature is summarized in Fig. 1. Researchers have studied these systems under different fluid flow conditions such as no flow, laminar and turbulent flows. Another classification of studies is based on the solar concentration ratio (SCR), viz., low ($< 25$), medium (25-50) and high ($> 50$) [see Fig. 1(a)]. Research studies can also be categorized based on the material (metallic and carbon-based nanostructures) and the orders of magnitude of the volume fraction of nanoparticles employed [see Fig. 1(b)].

Most of the reported modeling frameworks are either over-simplified and/or are not robust enough to tackle scales of design and operating parameters (see Fig. 1). For instance, majority of the reported works have not considered the radiation exchange within the fluid layers and between the enveloping surfaces; conjugate heat transfer between various receiver elements; temperature dependence of radiative parameters - this can lead to unrealistic results, particularly at high temperatures and high flux conditions [see Fig. 1(b)].

The present work serves to develop a comprehensive and mechanistic theoretical framework that is robust enough to account for coupled transport phenomena (within the fluid and between the enveloping surfaces) as well as conjugate heat transfer at the fluid-solid interface. Further, attempt has been made to include various receiver design configurations, orders of magnitude of SCRs and nanoparticle volume fractions in the laminar flow regime. Moreover, we have also modeled equivalent surface-absorption based receiver (SAR) designs; and have subsequently compared their performance characteristics with their corresponding volumetric absorption based counterparts.

Finally, to get the 'big picture' we have tried to delineate distinct range of values of design and operational parameters for which the nanofluid based volumetric absorption solar thermal platforms could prove to be beneficial.

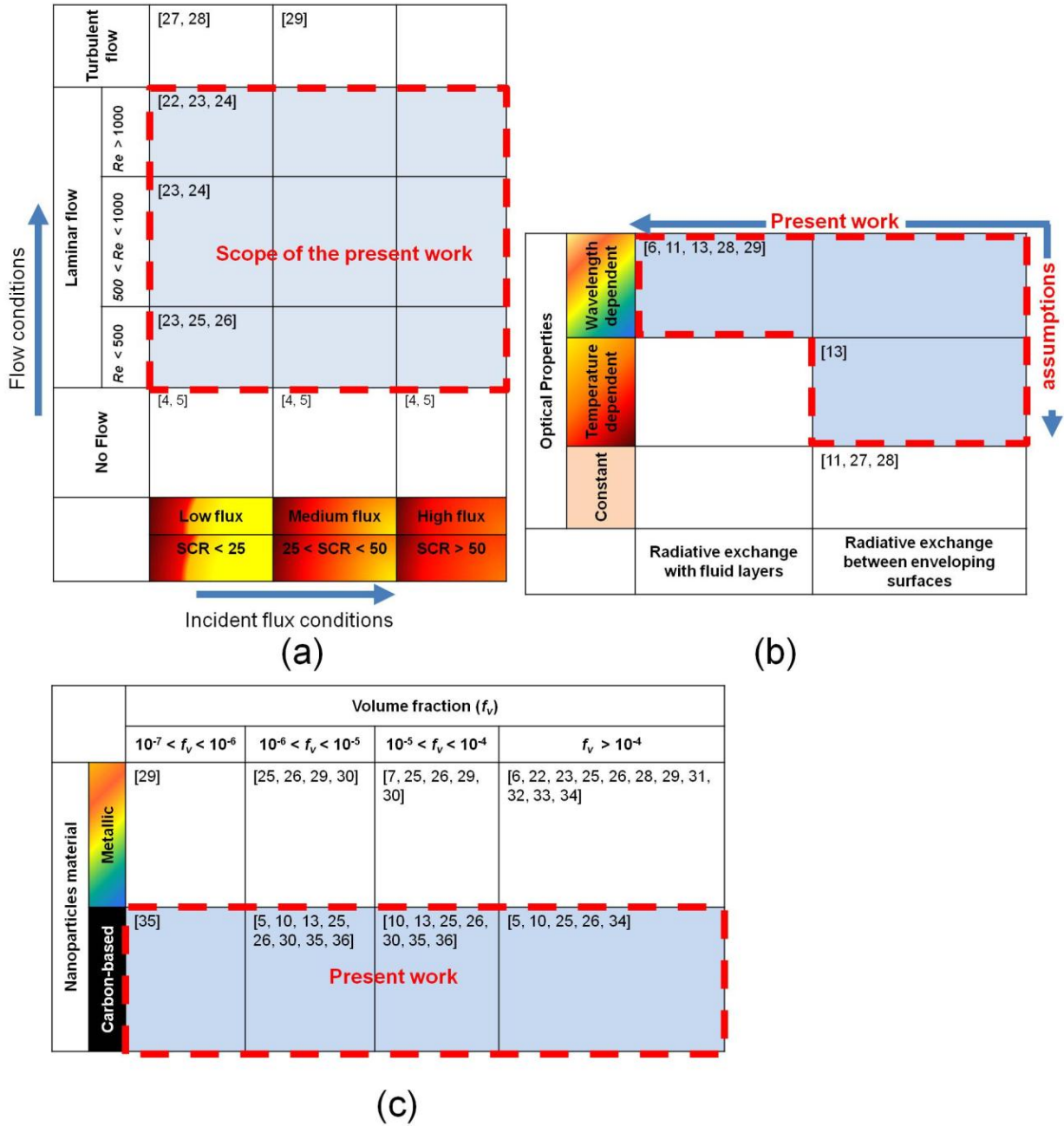

Fig. 1 Modeling of volumetric absorption receivers: (a) selected reported works for various flow conditions and SCRs, (b) modeling assumptions in relation to the optical properties and radiation exchange, and (c) selected reported works relevant to the nanoparticle materials and volume fractions.

## 2. Concept of volumetric absorption and design considerations
### 2.1 Basic concept

***2.1.1 Efficient photo-thermal energy conversion:*** Although, conventional heat transfer fluids (such as water, oil; termed as basefluids) are essentially transparent in the solar irradiance wavelength band; dispersing nanoparticles into these fluids has an enormous impact on their optical properties - allowing them to capture incident solar radiation which otherwise gets transmitted through these pristine basefluids [16, 37, 38].

Plasmonic heating of nanostructures (photo-thermal energy conversion process) involves the interaction between light and matter - the dimensions of matter being much smaller compared to the wavelength of light which irradiates it. As electromagnetic radiation (sunlight) impinges on nanostructures, it couples with the electron density at the surface of the particles [known as localized surface plasmons (LSPs)] generating heat via non-radiative decay mechanism [39]. The magnitude of the heat generated depends on the frequencies of the incoming radiation and the LSP frequencies of the nanoparticles which in turn depends on their material, shape, size and the surrounding medium [13,16]. In the backdrop of the aforementioned facts; the nanoparticles material, size, shape; basefluid should be carefully selected to ensure efficient photo-thermal conversion at low nanoparticle volume fractions.

***2.1.2 Mitigating thermal losses:*** As pointed out earlier, we can engineer solar absorbing nanoparticle dispersions through careful control of nanostructure morphology as well as the surrounding media (basefluid); however; given the fact that the nanoparticle dispersions are inherently good radiators of heat in the infrared region (due to intra-molecular vibrations), necessitates the use of enveloping surfaces to minimize thermal losses [13, 16]. If the operating temperatures are not too high, convective losses are predominant; then glass could be effective enough as the enveloping surface. However, at high temperatures, radiative losses are predominant; therefore transparent heat mirrors (which allow the sunlight to pass through but reflect the infrared radiation back to the nanofluid) could be used as the enveloping surface [11, 13, 40].

**2.2 Constructional details**

Figure 2 (a) - (c) show volumetric absorption based receiver design configurations. For comparison purposes, corresponding surface absorption based receiver designs have also been shown in Fig. 2 (d) - (e).

In all the VAR variants, the nanofluid is made to flow through a rectangular conduit having top surfaces such that it allows the solar irradiance to pass through and directly interact with the working fluid (nanofluid) [see Fig. 2(a) - 2(c)]. Whereas, in all the SAR variants, the working fluid (basefluid) is made to flow through a rectangular conduit having top surface such that it absorbs the incident radiation (i.e. either black or solar selective). Subsequently, the absorbed energy (in case of SAR) is then transferred from the absorbing surface to the basefluid through conduction and convection.

Furthermore, either heat mirror [Fig. 2(a) and 2(d)] or glass [Fig. 2(b) and 2(e))] forms the envelope else there is no cover and the conduit is exposed to the ambient conditions [Fig. 2(c) and 2(f)].

Although, all the aforementioned receiver configurations are distinct (in terms of constructional details and the relative importance of heat transfer mechanisms involved); the common design philosophy that transcends across these receivers is 'maximizing solar energy absorption and minimizing thermal losses'.

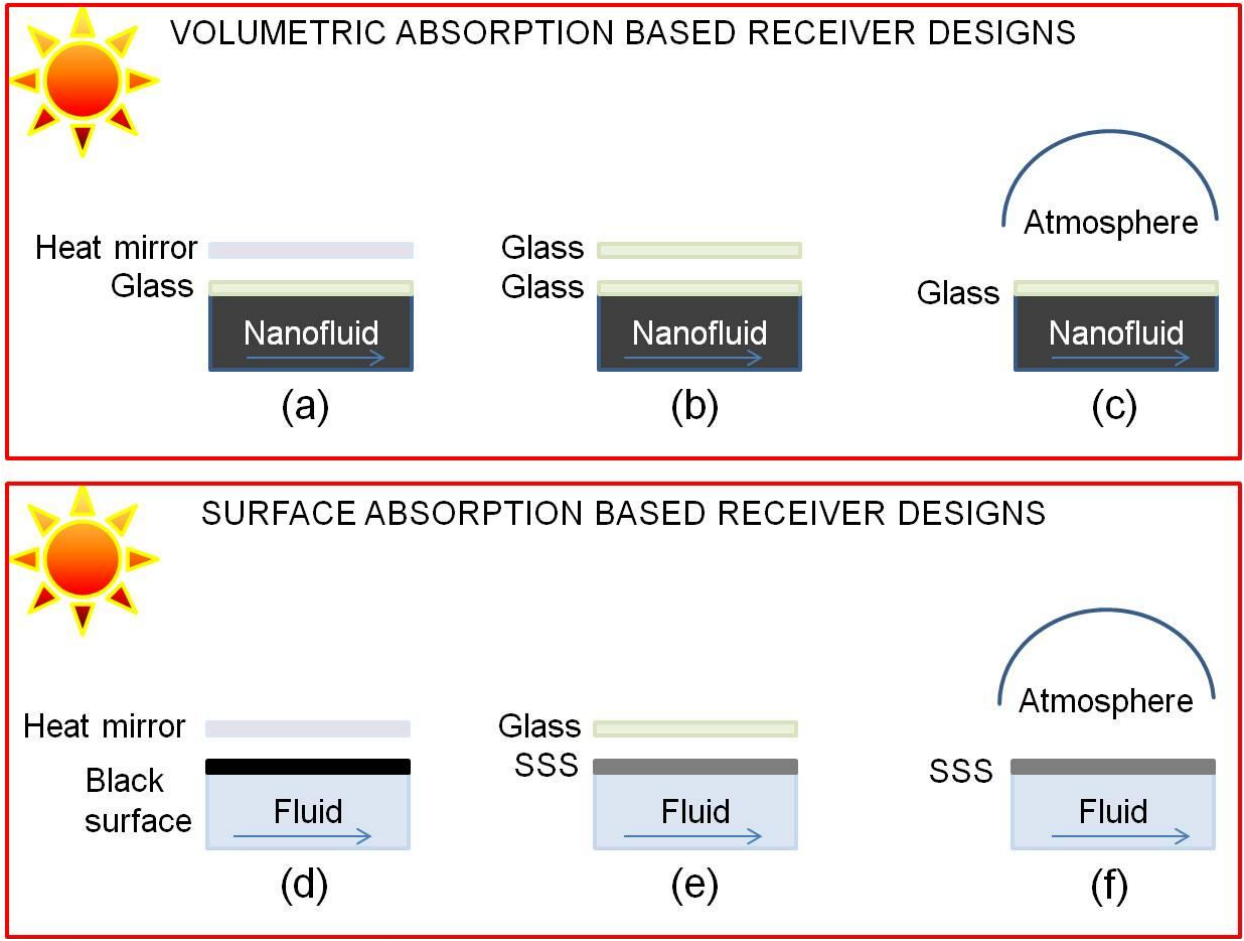

Fig. 2 Schematic showing VAR designs (a) glass - heat mirror (G-HM), (b) glass - glass (G-G), (c) glass - atmosphere (G-ATM), and SAR designs (d) black surface - heat mirror (BS-HM), (e) solar selective surface - glass (SSS-G), and solar selective surface - atmosphere (SSS-ATM).

## 3. Theoretical modeling framework
### 3.1 Spectral optical properties of constituent elements
Venn diagram detailing broad optical characteristics of constituent optical elements' viz., nanofluid, solar selective surface, black surface, basefluid, heat mirror, and glass is shown in Fig. 3(a). Optical elements could be broadly categorized into 'enveloping surfaces' and 'solar energy absorbing elements' as described below.

*3.1.1 Enveloping surfaces (glass and heat mirror):* Enveloping surfaces, whether for VARs or SARs should be highly transparent to the incident sunlight (short wavelength radiations) so that maximum amount of sunlight is able to reach the absorbing medium/surface [see Fig. 3(d)]. Further, these have inherently high absorptivity (glass) or reflectivity (heat mirror) values in the long wavelength infrared region [see Fig. 3(c)]. This lends them to be effective in mitigating thermal losses. The spectral transmissivity values for glass have been calculated (detailed in appendix A) utilizing data from Ref. [41]. The transmissivity values for heat mirrors have been taken from Ref. [13].

One can see that both glass and heat mirror have high transmissivity values at short wavelengths. However, at wavelength beyond 1$\mu$m (the cut-off wavelength in the present case), the transmissivity of heat mirror falls rapidly. This leads to a considerably lesser solar weighted transmissivity (defined by Eq. 1) for heat mirrors ($\tau_{sw}$ = 0.872) than for glass ($\tau_{sw}$ = 0.978); allowing more solar radiation to pass through in systems with a glass cover. However, the reduction in transmissivity values results due to corresponding increase in reflectivity. Therefore, trade-offs need to be made at different receiver operating temperatures (by tuning the value of cut-off wavelength) to ensure optimum performance.

$$\tau_{sw} = \frac{\int_{\lambda=0.3\mu m}^{\lambda=30\mu m} \tau_\lambda S_\lambda d\lambda}{\int_{\lambda=0.3\mu m}^{\lambda=30\mu m} S_\lambda d\lambda} \approx \frac{\sum_{\lambda=0.3\mu m}^{\lambda=30\mu m} \tau_\lambda S_\lambda d\lambda}{\sum_{\lambda=0.3\mu m}^{\lambda=30\mu m} S_\lambda d\lambda}, \quad (1)$$

where $S_\lambda$ is the spectral solar irradiance (AM 1.5) and $\tau_\lambda$ is the corresponding spectral transmissivity value of the enveloping surface (glass/heat mirror)

*3.1.2 Solar energy absorbing elements (nanofluid, solar selective surface, black surface):* These are characterized by high values of extinction coefficients (calculation detailed in appendix A)/absorptivity in the short wavelength solar irradiance region to ensure efficient photo-thermal energy conversion. Additionally, solar selective surfaces in particular are engineered to have low emissivity in the long wavelength infrared wavelength band to ensure low radiative losses. However, both nanofluids and black surfaces have high emissivity in the infrared region; therefore require enveloping surfaces [particularly heat mirrors, see Fig. 2(a) and 2(d)] to mitigate radiative losses. The emission spectra of black surface at temperatures 500K, 1000K and 1500K is also laid out in the graph showing the wavelengths at which peak emission occurs in each instance. The peak shifts towards the left as the temperature rises [see Fig. 3(c)]. Any combination of optical elements which ensures high photo-thermal energy conversion in conjunction with low thermal losses is suitable for solar thermal applications.

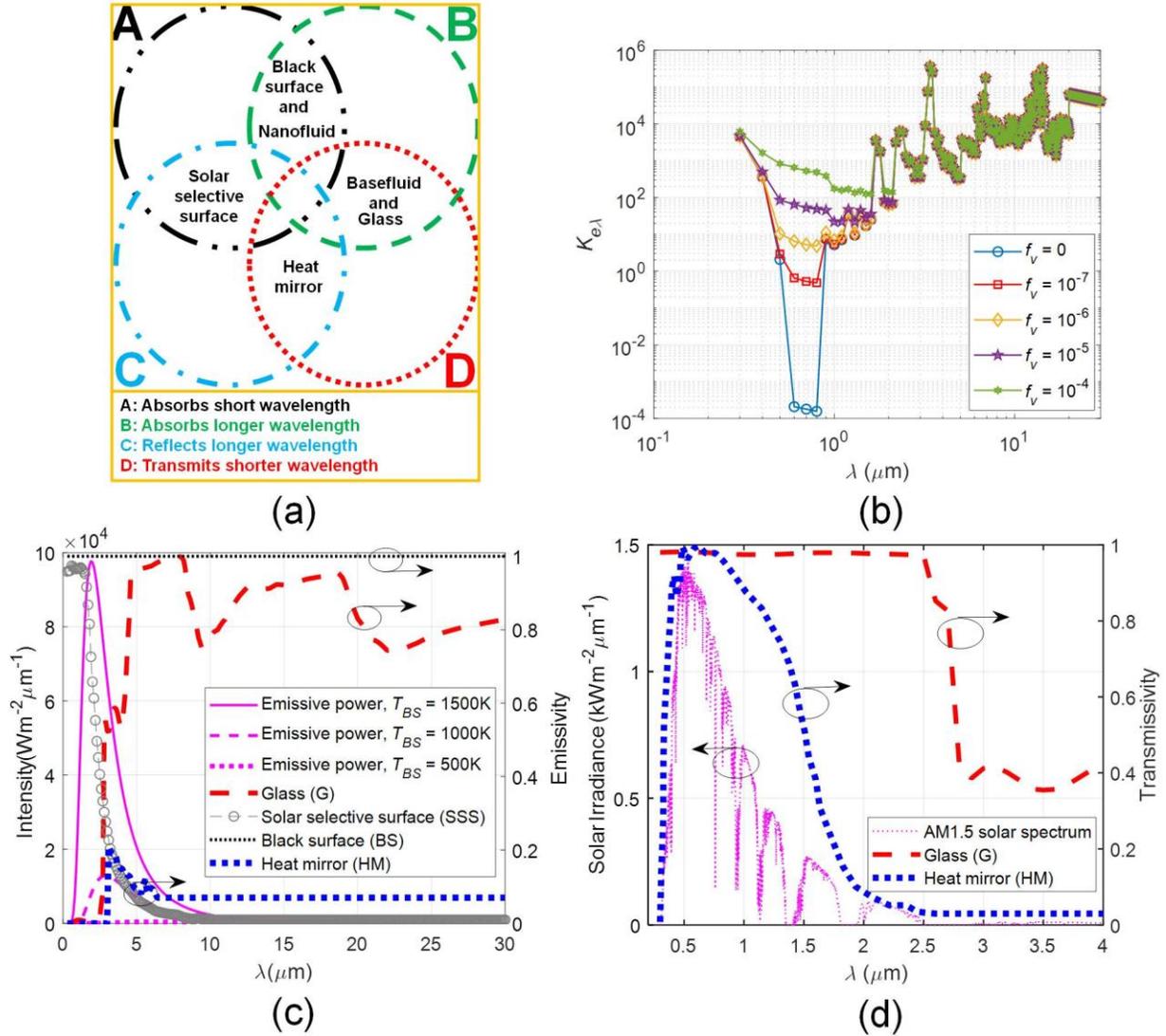

Fig. 3 (a) Venn diagram showing typical optical property characteristics of receiver constituent materials, (b) spectral extinction coefficients as a function of nanoparticle volume fraction, (c) spectral emissive power at various black body temperatures and spectral emissivity of various optical surfaces, and (d) AM 1.5 solar spectrum and spectral transmissivity of glass and heat mirror.

**3.2 Modeling heat transfer mechanisms**
Once we know the spectral optical properties of various constituent elements, the next step is to model heat transfer mechanisms involved in these systems.
Figure 4 shows the schematic detailing the heat transfer mechanisms involved in VARs and SARs.

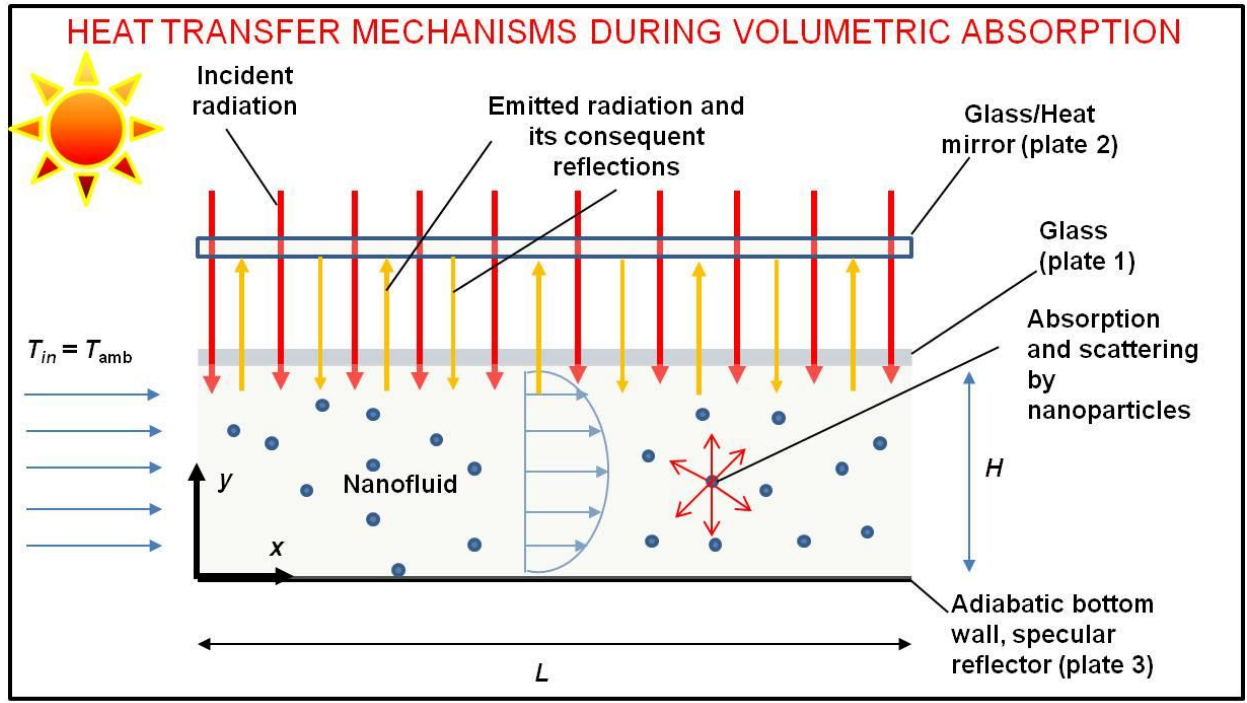

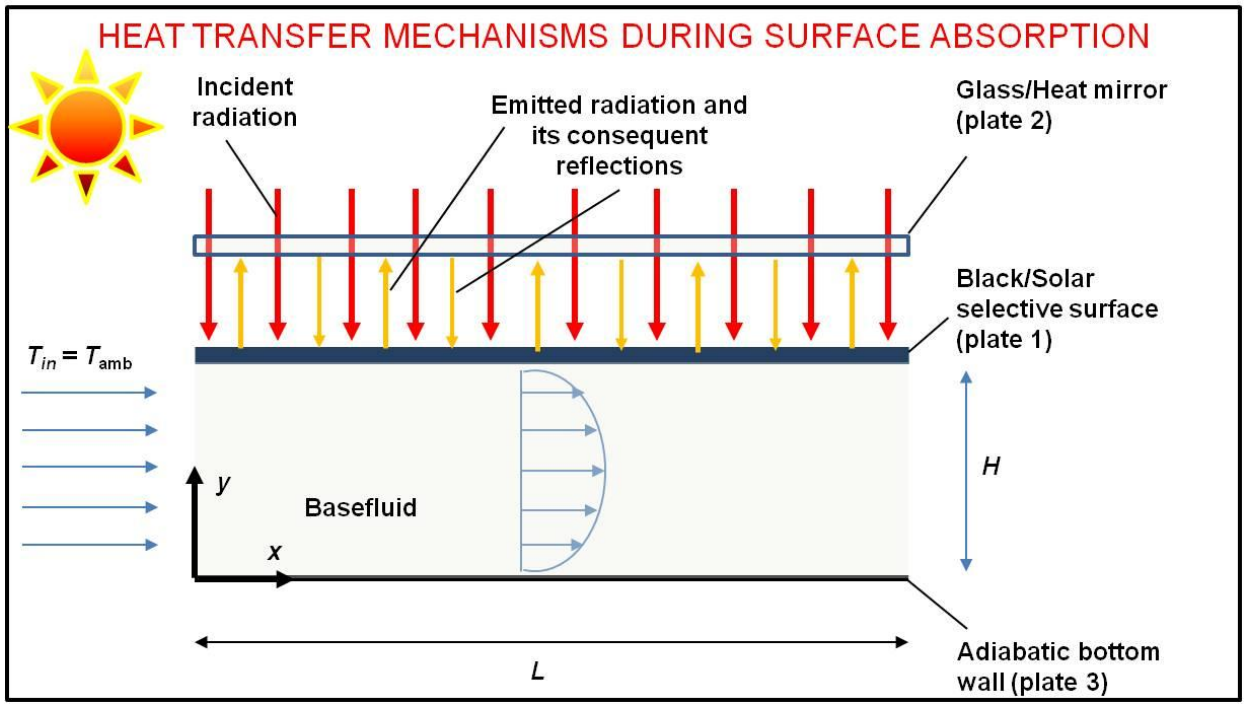

Fig. 4 Schematic showing the heat transfer mechanisms involved in (a) VARs, and (b) SARs

The developed models are based on certain simplifying assumptions which transcends across both VARs and SARs (described in subsection 3.2.1)). Moreover, certain aspects such as radiation exchange between the covers are common to both VARs and SARs; and have been described in detail in subsection 3.2.2. However, there are certain aspects which are specific to the nature of absorption mechanisms and subsequent heat transfers involved (i.e., characteristics of VARs and SARs); therefore have been treated separately in detail.

*3.2.1 Underlying assumptions:* Following are the modeling assumptions which are common to both VARs and SARs:

1) The flow inside the channel is assumed to be fully developed with a parabolic velocity profile given by Eq. (2)

$$u_y = 6u_{av}\left[\frac{y}{H} - \left(\frac{y}{H}\right)^2\right], \quad (2)$$

where

$$u_{av} = \frac{\text{Re}.\mu_f}{D_f.(2H)} \quad (3)$$

The y-component of velocity is assumed to be zero.

2) Heat transfer by conduction in the *x*-direction has been ignored as convection from the moving fluid is the predominant mode of heat transfer.

3) There is nearly vacuum between the two plates, i.e., no convective heat transfer between the bottom and the top plate.

4) Heat transfer coefficient between the casing and the atmosphere is assumed to be $10\text{Wm}^{-2}\text{K}^{-1}$ [42].

*3.2.2 Radiation exchange between two parallel plates (the top surface of channel and the casing that encloses it):* The net heat lost by the top plate of the conduit (referred to as plate 1) and the heat gained by the cover plate (referred to as plate 2) where one or both are semi-transparent is explained in this section. A system of two plates with known spectral optical properties is considered. As a result of multiple reflections taking place between the two plates, and the fact that the radiation is spectral in nature and two-way coupling exists between the optical behavior and plate temperatures; we need to define parameters such as effective emissivity, absorptivity, reflectivity, and transmissivity to quantitatively determine the overall optical characteristics of the interacting plates.

*Effective emissivity:* Effective emissivity [mathematically defined by Eq. (4)] is the measure of emission from a surface relative to a corresponding black body at same temperature. It is a material property and also a function of temperature. Figure 5 shows the effective emissivity values as a function of temperature for glass, heat mirror and solar selective surface.

$$\varepsilon_{eff} = \frac{\int_{0.3\mu m}^{30\mu m}(1-\rho_\lambda-\tau_\lambda)e_{b,T,\lambda}d\lambda}{\int_{0.3\mu m}^{30\mu m}e_{b,T,\lambda}d\lambda} = \frac{\sum_{0.3\mu m}^{30\mu m}(1-\rho_\lambda-\tau_\lambda)e_{b,T,\lambda}d\lambda}{\sum_{0.3\mu m}^{30\mu m}e_{b,T,\lambda}d\lambda} = \frac{\sum_{\lambda=0.3\mu m}^{\lambda=30\mu m}\varepsilon_{\lambda,T}e_{b,\lambda,T}d\lambda}{\sum_{\lambda=0.3\mu m}^{\lambda=30\mu m}e_{b,\lambda,T}d\lambda} \quad (4)$$

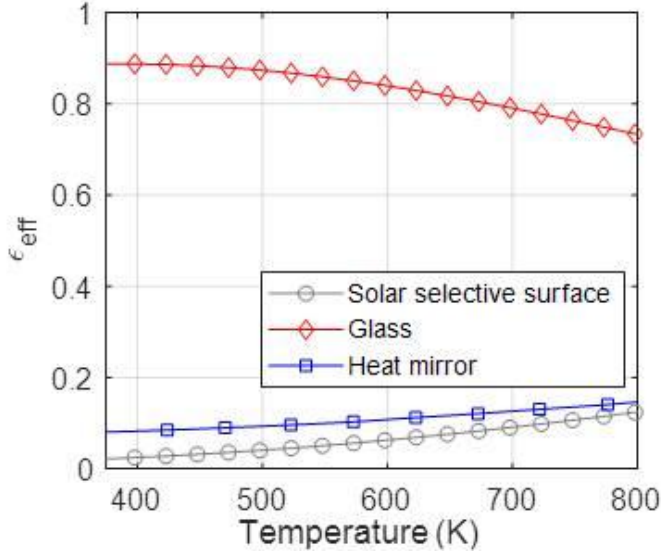

Fig. 5 Effective emissivity as a function of temperature for various optical elements, viz., solar selective surface, glass, and heat mirror.

*Effective absorptivity, reflectivity, and transmissivity:* Parameters viz., effective absorptivity [defined by Eqs. (5) and (6)]), reflectivity [defined by Eqs.(7) and (8)], and transmissivity [defined by Eqs.(9) and (10)] are not material properties and depend on the irradiation spectra. Since the magnitude and spectra of the radiation falling on any of the plates varies with each reflection, these keep on changing as well.

$$\alpha_{1j,eff} = \frac{\sum_{\lambda=0.3\mu m}^{\lambda=30\mu m} \varepsilon_{\lambda,1} e_{b,\lambda,T1} \rho_{\lambda,2}^{j} \rho_{\lambda,1}^{j-1} \alpha_{\lambda,1} d\lambda}{\sum_{\lambda=0.3\mu m}^{\lambda=30\mu m} \varepsilon_{\lambda,1} e_{b,\lambda,T1} \rho_{\lambda,2}^{j} \rho_{\lambda,1}^{j-1} d\lambda} \ , \ \alpha_{2j,eff} = \frac{\sum_{\lambda=0.3\mu m}^{\lambda=30\mu m} \varepsilon_{\lambda,1} e_{b,\lambda,T1} \rho_{\lambda,1}^{j-1} \rho_{\lambda,2}^{j-1} \alpha_{\lambda,2} d\lambda}{\sum_{\lambda=0.3\mu m}^{\lambda=30\mu m} \varepsilon_{\lambda,1} e_{b,\lambda,T1} \rho_{\lambda,1}^{j-1} \rho_{\lambda,2}^{j-1} d\lambda} \ , \quad (5)$$

$$\alpha'_{1j,eff} = \frac{\sum_{\lambda=0.3\mu m}^{\lambda=30\mu m} \varepsilon_{\lambda,2} e_{b,\lambda,T2} \rho_{\lambda,2}^{j-1} \rho_{\lambda,1}^{j-1} \alpha_{\lambda,1} d\lambda}{\sum_{\lambda=0.3\mu m}^{\lambda=30\mu m} \varepsilon_{\lambda,2} e_{b,\lambda,T2} \rho_{\lambda,2}^{j-1} \rho_{\lambda,1}^{j-1} d\lambda} \ , \ \alpha'_{2j,eff} = \frac{\sum_{\lambda=0.3\mu m}^{\lambda=30\mu m} \varepsilon_{\lambda,2} e_{b,\lambda,T2} \rho_{\lambda,1}^{j} \rho_{\lambda,2}^{j-1} \alpha_{\lambda,2} d\lambda}{\sum_{\lambda=0.3\mu m}^{\lambda=30\mu m} \varepsilon_{\lambda,2} e_{b,\lambda,T2} \rho_{\lambda,1}^{j} \rho_{\lambda,2}^{j-1} d\lambda} \ , \quad (6)$$

$$\rho_{1j,eff} = \frac{\sum_{\lambda=0.3\mu m}^{\lambda=30\mu m} \varepsilon_{\lambda,1} e_{b,\lambda,T1} \rho_{\lambda,2}^{j} \rho_{\lambda,1}^{j} d\lambda}{\sum_{\lambda=0.3\mu m}^{\lambda=30\mu m} \varepsilon_{\lambda,1} e_{b,\lambda,T1} \rho_{\lambda,2}^{j} \rho_{\lambda,1}^{j-1} d\lambda} \ , \ \rho_{2j,eff} = \frac{\sum_{\lambda=0.3\mu m}^{\lambda=30\mu m} \varepsilon_{\lambda,1} e_{b,\lambda,T1} \rho_{\lambda,2}^{j} \rho_{\lambda,1}^{j-1} d\lambda}{\sum_{\lambda=0.3\mu m}^{\lambda=30\mu m} \varepsilon_{\lambda,1} e_{b,\lambda,T1} \rho_{\lambda,2}^{j} \rho_{\lambda,1}^{j-1} d\lambda} \ , \quad (7)$$

$$\rho'_{1j,eff} = \frac{\sum_{\lambda=0.3\mu m}^{\lambda=30\mu m} \varepsilon_{\lambda,2} e_{b,\lambda,T2} \rho_{\lambda,2}^{j-1} \rho_{\lambda,1}^{j} d\lambda}{\sum_{\lambda=0.3\mu m}^{\lambda=30\mu m} \varepsilon_{\lambda,2} e_{b,\lambda,T2} \rho_{\lambda,2}^{j-1} \rho_{\lambda,1}^{j-1} d\lambda}, \quad \rho'_{2j,eff} = \frac{\sum_{\lambda=0.3\mu m}^{\lambda=30\mu m} \varepsilon_{\lambda,2} e_{b,\lambda,T2} \rho_{\lambda,2}^{j} \rho_{\lambda,1}^{j} d\lambda}{\sum_{\lambda=0.3\mu m}^{\lambda=30\mu m} \varepsilon_{\lambda,2} e_{b,\lambda,T2} \rho_{\lambda,2}^{j-1} \rho_{\lambda,1}^{j} d\lambda}, \tag{8}$$

$$\tau'_{1j,eff} = \frac{\sum_{\lambda=0.3\mu m}^{\lambda=30\mu m} \varepsilon_{\lambda,1} e_{b,\lambda,T1} \rho_{\lambda,2}^{j} \rho_{\lambda,1}^{j-1} \tau_{\lambda,1} d\lambda}{\sum_{\lambda=0.3\mu m}^{\lambda=30\mu m} \varepsilon_{\lambda,1} e_{b,\lambda,T1} \rho_{\lambda,2}^{j} \rho_{\lambda,1}^{j-1} d\lambda}, \quad \tau'_{2j,eff} = \frac{\sum_{\lambda=0.3\mu m}^{\lambda=30\mu m} \varepsilon_{\lambda,1} e_{b,\lambda,T1} \rho_{\lambda,1}^{j-1} \rho_{\lambda,2}^{j-1} \tau_{\lambda,2} d\lambda}{\sum_{\lambda=0.3\mu m}^{\lambda=30\mu m} \varepsilon_{\lambda,1} e_{b,\lambda,T1} \rho_{\lambda,1}^{j-1} \rho_{\lambda,2}^{j-1} d\lambda}, \tag{9}$$

$$\tau'_{1j,eff} = \frac{\sum_{\lambda=0.3\mu m}^{\lambda=30\mu m} \varepsilon_{\lambda,2} e_{b,\lambda,T2} \rho_{\lambda,2}^{j-1} \rho_{\lambda,1}^{j-1} \tau_{\lambda,1} d\lambda}{\sum_{\lambda=0.3\mu m}^{\lambda=30\mu m} \varepsilon_{\lambda,2} e_{b,\lambda,T2} \rho_{\lambda,2}^{j-1} \rho_{\lambda,1}^{j-1} d\lambda}, \quad \tau'_{2j,eff} = \frac{\sum_{\lambda=0.3\mu m}^{\lambda=30\mu m} \varepsilon_{\lambda,2} e_{b,\lambda,T2} \rho_{\lambda,1}^{j} \rho_{\lambda,2}^{j-1} \tau_{\lambda,2} d\lambda}{\sum_{\lambda=0.3\mu m}^{\lambda=30\mu m} \varepsilon_{\lambda,2} e_{b,\lambda,T2} \rho_{\lambda,1}^{j} \rho_{\lambda,2}^{j-1} d\lambda}, \tag{10}$$

where '$j$' is the number of reflections.

Also, Eq. (11) gives the effective absorptivity of atmosphere

$$\alpha_{3,eff} = \frac{\sum_{\lambda=0.3\mu m}^{\lambda=30\mu m} \alpha_{2,\lambda,T_{amb}} e_{b,\lambda,T_{amb}} d\lambda}{\sum_{\lambda=0.3\mu m}^{\lambda=30\mu m} e_{b,\lambda,T_{amb}} d\lambda} \tag{11}$$

*Heat lost by plate 1:* In case of radiation exchange between semitransparent plates (unlike opaque surfaces), the net heat lost by one plate does not equal the net heat gained by the second plate. Figure 5 shows the interaction of two semitransparent or one semitransparent and one opaque plates.

$$Q_{loss,1} = \varepsilon_{1,eff} \sigma T_{1,avg}^{4}[1 - \rho_{21,eff}\alpha_{11,eff} - \rho_{21,eff}\rho_{22,eff}\rho_{11,eff}\alpha_{12,eff} - \rho_{21,eff}\rho_{22,eff}\rho_{23,eff}\rho_{11,eff}\rho_{12,eff}\alpha_{13,eff} - \ldots] - \varepsilon_{2,eff}\sigma T_{2,avg}^{4}[\alpha'_{11,eff} + \rho'_{11,eff}\rho'_{21,eff}\alpha'_{12,eff} + \rho'_{11,eff}\rho'_{12,eff}\rho'_{21,eff}\rho'_{22,eff}\alpha'_{13,eff} + \ldots] \tag{12}$$

*Heat gained by plate 2:* The net heat gained by plate 2 is given by Eq. (13) as

$$Q_{gain,2} = \varepsilon_{2,eff}\sigma T_{2,avg}^{4}[-1 + \rho'_{11,eff}\alpha'_{21,eff} + \rho'_{11,eff}\rho'_{12,eff}\rho'_{21,eff}\alpha'_{21,eff} + \rho'_{11,eff}\rho'_{12,eff}\rho'_{13,eff}\rho'_{21,eff}\rho'_{22,eff}\alpha'_{23,eff} + \ldots] + \varepsilon_{1,eff}\sigma T_{1,avg}^{4}[\alpha_{21,eff} + \rho_{11,eff}\rho_{21,eff}\alpha_{22,eff} + \rho_{11,eff}\rho_{12,eff}\rho_{21,eff}\rho_{22,eff}\alpha_{23,eff}] + \alpha_{3,eff}\sigma T_{amb}^{4} \tag{13}$$

Similarly, one can find the value of energy transmitted through to the nanofluid ($Q_{\text{trans}\to\text{nf}}$) flowing underneath the lower plate by summing flux at points $l_1$, $m_1$, $n_1$….and $s_2$, $t_2$, $u_2$, $v_2$ ….as given in Fig. 6. In order to find the temperature of top cover (plate 2) when the temperature of the top plate of conduit (plate 1) is known, we invoke a relationship for energy balance by equating the energy gained by the top cover (plate 2) from the plate (plate 1) with the energy lost by the cover (plate 2) due radiation and convection losses as given by Eq. (14)

$$\varepsilon_{2,eff}\sigma T_{2,avg}^{4} + \alpha_{3,eff}\sigma T_{amb}^{4} + h_{wind}(T_{2,avg} - T_{amb}) = \varepsilon_{2,eff}\sigma T_{2,avg}^{4}[-1 + \rho'_{11,eff}\alpha'_{21,eff} + \rho'_{11,eff}\rho'_{12,eff}\rho'_{21,eff}\alpha'_{21,eff} + \ldots] +$$
$$\varepsilon_{1,eff}\sigma T_{1,avg}^{4}[\alpha_{21,eff} + \rho_{11,eff}\rho_{21,eff}\alpha_{22,eff} + \rho_{11,eff}\rho_{12,eff}\rho_{21,eff}\rho_{22,eff}\alpha_{23,eff}]$$
(14)

On solving Eq. (14), if the value of $T_{1,avg}$ (plate 1 average temperature) is known, the value of $T_{2,avg}$ (plate 2 average temperature) can be obtained. Figure 7 unveils the relationship between the top and bottom plate temperatures for various combinations of plate materials.

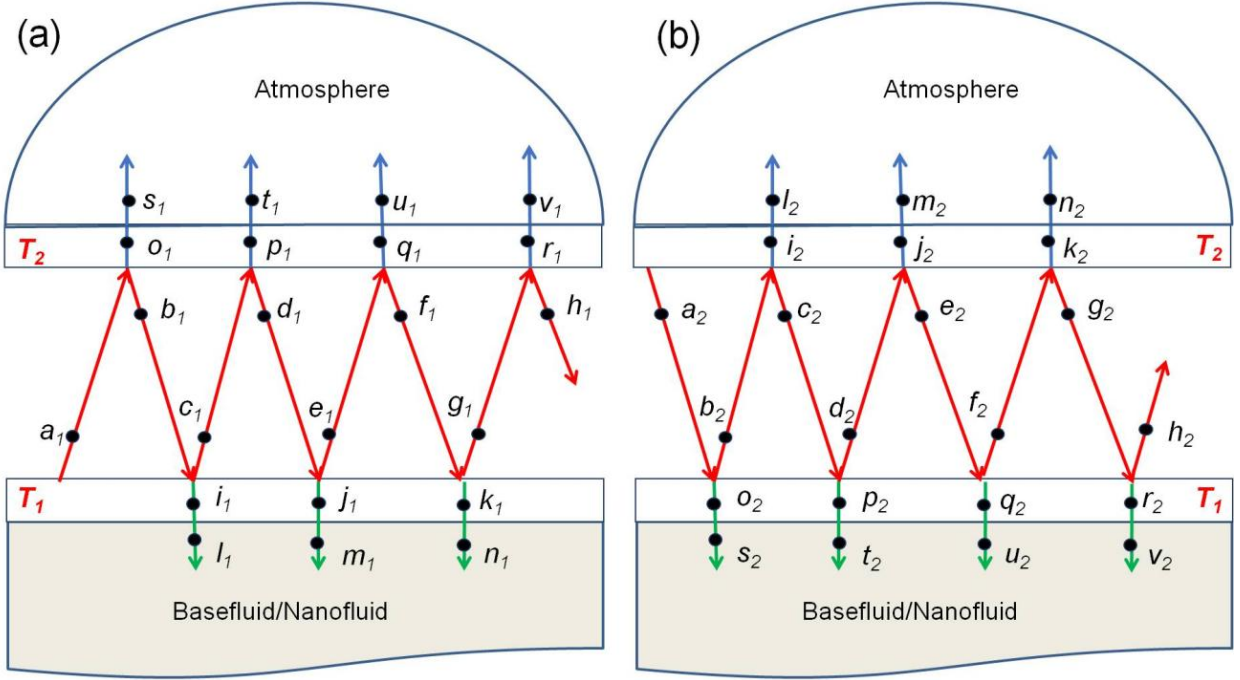

| Location | Magnitude of flux | Location | Magnitude of flux |
|---|---|---|---|
| $a_1$ | $\varepsilon_{1,eff}\sigma T_1^4$ | $a_2$ | $\varepsilon_{2,eff}\sigma T_2^4$ |
| $b_1$ | $\varepsilon_{1,eff}\sigma T_1^4 \rho_{21,eff}$ | $b_2$ | $\varepsilon_{2,eff}\sigma T_2^4 \rho'_{11,eff}$ |
| $c_1$ | $\varepsilon_{1,eff}\sigma T_1^4 \rho_{21,eff}\rho_{11,eff}$ | $c_2$ | $\varepsilon_{2,eff}\sigma T_2^4 \rho'_{11,eff}\rho'_{21,eff}$ |
| $d_1$ | $\varepsilon_{1,eff}\sigma T_1^4 \rho_{21,eff}\rho_{22,eff}\rho_{11,eff}$ | $d_2$ | $\varepsilon_{2,eff}\sigma T_2^4 \rho'_{11,eff}\rho'_{12,eff}\rho'_{21,eff}$ |
| $e_1$ | $\varepsilon_{1,eff}\sigma T_1^4 \rho_{21,eff}\rho_{22,eff}\rho_{11,eff}\rho_{12,eff}$ | $e_2$ | $\varepsilon_{2,eff}\sigma T_2^4 \rho'_{11,eff}\rho'_{12,eff}\rho'_{21,eff}\rho'_{22,eff}$ |
| $f_1$ | $\varepsilon_{1,eff}\sigma T_1^4 \rho_{21,eff}\rho_{22,eff}\rho_{23,eff}\rho_{11,eff}\rho_{12,eff}$ | $f_2$ | $\varepsilon_{2,eff}\sigma T_2^4 \rho'_{11,eff}\rho'_{12,eff}\rho'_{13,eff}\rho'_{21,eff}\rho'_{22,eff}$ |
| $g_1$ | $\varepsilon_{1,eff}\sigma T_1^4 \rho_{21,eff}\rho_{22,eff}\rho_{23,eff}\rho_{11,eff}\rho_{12,eff}\rho_{13,eff}$ | $g_2$ | $\varepsilon_{2,eff}\sigma T_2^4 \rho'_{11,eff}\rho'_{12,eff}\rho'_{13,eff}\rho'_{21,eff}\rho'_{22,eff}\rho'_{23,eff}$ |
| $h_1$ | $\varepsilon_{1,eff}\sigma T_1^4 \rho_{21,eff}\rho_{22,eff}\rho_{23,eff}\rho_{24,eff}\rho_{11,eff}\rho_{12,eff}\rho_{13,eff}$ | $h_2$ | $\varepsilon_{2,eff}\sigma T_2^4 \rho'_{11,eff}\rho'_{12,eff}\rho'_{13,eff}\rho'_{14,eff}\rho'_{21,eff}\rho'_{22,eff}\rho'_{23,eff}$ |
| $i_1$ | $\varepsilon_{1,eff}\sigma T_1^4 \rho_{21,eff}\alpha_{11,eff}$ | $i_2$ | $\varepsilon_{2,eff}\sigma T_2^4 \rho'_{11,eff}\alpha'_{21,eff}$ |
| $j_1$ | $\varepsilon_{1,eff}\sigma T_1^4 \rho_{21,eff}\rho_{22,eff}\rho_{11,eff}\alpha_{12,eff}$ | $j_2$ | $\varepsilon_{2,eff}\sigma T_2^4 \rho'_{11,eff}\rho'_{12,eff}\rho'_{21,eff}\alpha'_{22,eff}$ |
| $k_1$ | $\varepsilon_{1,eff}\sigma T_1^4 \rho_{21,eff}\rho_{22,eff}\rho_{23,eff}\rho_{11,eff}\rho_{12,eff}\alpha_{13,eff}$ | $k_2$ | $\varepsilon_{2,eff}\sigma T_2^4 \rho'_{11,eff}\rho'_{12,eff}\rho'_{13,eff}\rho'_{21,eff}\rho'_{22,eff}\alpha'_{23,eff}$ |
| $l_1$ | $\varepsilon_{1,eff}\sigma T_1^4 \rho_{21,eff}\tau_{11,eff}$ | $l_2$ | $\varepsilon_{2,eff}\sigma T_2^4 \rho'_{11,eff}\tau'_{21,eff}$ |
| $m_1$ | $\varepsilon_{1,eff}\sigma T_1^4 \rho_{21,eff}\rho_{22,eff}\rho_{11,eff}\tau_{12,eff}$ | $m_2$ | $\varepsilon_{2,eff}\sigma T_2^4 \rho'_{11,eff}\rho'_{12,eff}\rho'_{21,eff}\tau'_{22,eff}$ |
| $n_1$ | $\varepsilon_{1,eff}\sigma T_1^4 \rho_{21,eff}\rho_{22,eff}\rho_{23,eff}\rho_{11,eff}\rho_{12,eff}\tau_{13,eff}$ | $n_2$ | $\varepsilon_{2,eff}\sigma T_2^4 \rho'_{11,eff}\rho'_{12,eff}\rho'_{13,eff}\rho'_{21,eff}\rho'_{22,eff}\tau'_{23,eff}$ |
| $o_1$ | $\varepsilon_{1,eff}\sigma T_1^4 \alpha_{21,eff}$ | $o_2$ | $\varepsilon_{2,eff}\sigma T_2^4 \alpha'_{11,eff}$ |
| $p_1$ | $\varepsilon_{1,eff}\sigma T_1^4 \rho_{21,eff}\rho_{11,eff}\alpha_{22,eff}$ | $p_2$ | $\varepsilon_{2,eff}\sigma T_2^4 \rho'_{11,eff}\rho'_{21,eff}\alpha'_{12,eff}$ |
| $q_1$ | $\varepsilon_{1,eff}\sigma T_1^4 \rho_{21,eff}\rho_{22,eff}\rho_{11,eff}\rho_{12,eff}\alpha_{23,eff}$ | $q_2$ | $\varepsilon_{2,eff}\sigma T_2^4 \rho'_{11,eff}\rho'_{12,eff}\rho'_{21,eff}\rho'_{22,eff}\alpha'_{13,eff}$ |
| $r_1$ | $\varepsilon_{1,eff}\sigma T_1^4 \rho_{21,eff}\rho_{22,eff}\rho_{23,eff}\rho_{11,eff}\rho_{12,eff}\rho_{13,eff}\alpha_{24,eff}$ | $r_2$ | $\varepsilon_{2,eff}\sigma T_2^4 \rho'_{11,eff}\rho'_{12,eff}\rho'_{13,eff}\rho'_{21,eff}\rho'_{22,eff}\rho'_{23,eff}\alpha'_{14,eff}$ |
| $s_1$ | $\varepsilon_{1,eff}\sigma T_1^4 \tau_{21,eff}$ | $s_2$ | $\varepsilon_{2,eff}\sigma T_2^4 \tau'_{11,eff}$ |
| $t_1$ | $\varepsilon_{1,eff}\sigma T_1^4 \rho_{21,eff}\rho_{11,eff}\tau_{22,eff}$ | $t_2$ | $\varepsilon_{2,eff}\sigma T_2^4 \rho'_{11,eff}\rho'_{21,eff}\tau'_{12,eff}$ |
| $u_1$ | $\varepsilon_{1,eff}\sigma T_1^4 \rho_{21,eff}\rho_{22,eff}\rho_{11,eff}\rho_{12,eff}\tau_{23,eff}$ | $u_2$ | $\varepsilon_{2,eff}\sigma T_2^4 \rho'_{11,eff}\rho'_{12,eff}\rho'_{21,eff}\rho'_{22,eff}\tau'_{13,eff}$ |
| $v_1$ | $\varepsilon_{1,eff}\sigma T_1^4 \rho_{21,eff}\rho_{22,eff}\rho_{23,eff}\rho_{11,eff}\rho_{12,eff}\rho_{13,eff}\tau_{24,eff}$ | $v_2$ | $\varepsilon_{2,eff}\sigma T_2^4 \rho'_{11,eff}\rho'_{12,eff}\rho'_{13,eff}\rho'_{21,eff}\rho'_{22,eff}\rho'_{23,eff}\tau'_{14,eff}$ |

Fig. 6 Schematic showing radiation exchange between two parallel plates: ray tracing (and corresponding equations) of the energy emanating from the (a) bottom (plate 1), and (b) top plate (plate 2).

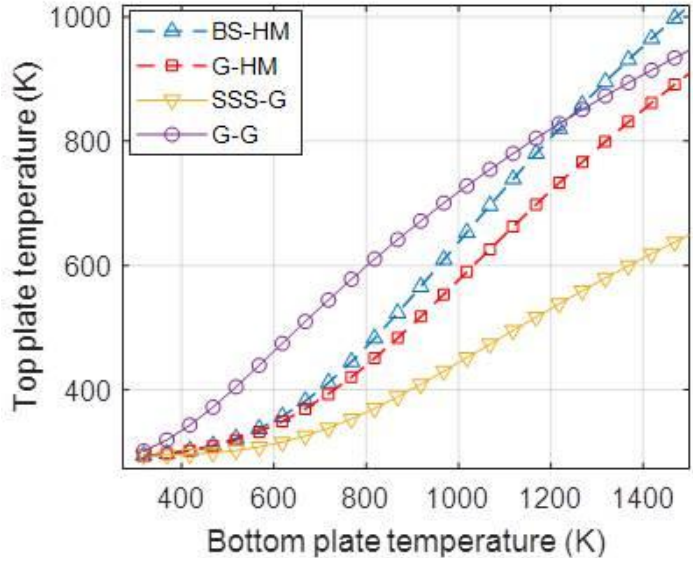

Fig. 7 Top plate temperature as a function of bottom plate temperature for various combinations of plate materials.

*3.2.3 Volumetric absorption based solar thermal systems:* The algorithm in Fig 8 gives a brief overview of the process of theoretical modeling pertinent to VARs.

On its way to the nanofluid, sunlight interacts with the two cover plates. The outer cover (plate 2) could be glass or heat mirror; whereas the plate in contact with the nanofluid is essentially glass (plate 1). The heat gain by plate 2 due to radiation exchange with plate 1 and the heat loss to the atmosphere (via convection and radiation) represents the overall energy balance for plate 2 and is given by Eq. (14). However, for plate 1; in addition to heat loss due to radiation exchange with plate 2, it is also in thermal contact with the nanofluid (i.e. conjugate heat transfer exists between plate 1 and nanofluid). Equations (15) – (18) describe the governing, initial and boundary conditions for the plate 1. It may be noted that the term $Q_{loss,\,1}$ is based on the temperatures of the two cover plates; which in turn requires solving Eq. (15) subject to initial and boundary conditions [Eqs. (16) – (18)]. Further, in order to solve Eq. (18), we need to solve RTE and the overall energy equations for the nanofluid.

$$\frac{\partial T}{\partial t} = \frac{k_1}{D_1 c_{p1}} \left[ \frac{\partial^2 T}{\partial y^2} \right] - \frac{1}{D_1 c_{p1}} \frac{\partial Q_{loss,1}}{\partial y} + \frac{1}{D_1 c_{p1}} \frac{\partial}{\partial y} \int \tau_{2\lambda} S_\lambda \left( 1 - e^{-K_{a\lambda,1}(H+t_1-y)} \right) d\lambda \qquad (15)$$

$$T(y)\big|_{t=0} = T_{amb} \qquad (16)$$

$$Q\big|_{y=H+t_1} = \int \tau_{2\lambda} S_\lambda d\lambda - Q_{loss,1} \qquad (17)$$

$$Q\big|_{y=H} = \frac{k_1}{D_1 c_{p1}}\left[\frac{\partial T}{\partial y}\right] + \int\left(1 - e^{-K_{a\lambda,1}(t_1)}\right)\left(q_{\lambda(H)}^{-} - q_{b\lambda[T_{y=H}]}^{+}\right)d\lambda \tag{18}$$

where $q_{b\lambda[T_{y=H}]}^{+} = 2\pi \int_0^1 I_{b\lambda[T_{y=H}]}^{+} \phi \, d\phi$

Within the nanofluid, the sunlight interacts primarily with nanoparticles through absorption and scattering mechanisms. In these processes, both the radiative (manifested in the form of emission and scattering phenomena) as well as non-radiative (manifested in the form of absorption phenomena) decay of incident sunlight takes place. The RTE provides the value of radiation intensity along the line of sight by accounting for the aforementioned three processes viz., absorption, emission and scattering. A numerical method for solving the RTE has been used to calculate the value of intensity at various points in the positive and negative directions. Once the values of intensity in the positive and negative directions are known across the fluid depth, one can calculate the value of the net heat flux leaving a control volume (and hence the divergence of radiative flux). The details of the discretization strategy along with procedure for solving RTE are provided in appendices B and C respectively.

Once RTE has been solved for initial temperature distribution and the value of divergence is known *'a priori'*, one can substitute this value of divergence into the energy equation as energy generation term and establish the values of temperature field at new time instant. Explicit form of finite difference formulation has been used to numerically solve the energy equation. With the solving of the energy equation, a new temperature distribution is obtained for the next time instant. Once again, radiation exchange between the plates (as detailed in subsection 3.2.2) and the RTE is solved and new values of divergence are calculated for input to the energy equation. This process gives a transient temperature distribution at different time intervals and it is repeated until steady state has been reached.

Equations (19) – (22) represent the governing, initial and boundary conditions for the nanofluid.

$$\frac{\partial T}{\partial t} = \frac{k_{nf}}{D_{nf} c_{p,nf}}\left[\frac{\partial^2 T}{\partial y^2}\right] - \frac{1}{D_{nf} c_{p,nf}}\frac{\partial Q_{rad,nf}}{\partial y} + \frac{1}{D_{nf} c_{p,nf}}\frac{\partial Q_{trans \to nf}}{\partial y} - u_y \frac{\partial T}{\partial x} \tag{19}$$

$$T(y)\big|_{t=0} = T_{amb} \tag{20}$$

$$Q\big|_{y=H} = \frac{k_{nf}}{D_{nf} c_{pnf}}\left[\frac{\partial T}{\partial y}\right] + \int q_{\lambda[y=H]}^{+} d\lambda \tag{21}$$

$$Q\big|_{y=0} = \int \rho_{\lambda 3} q_{\lambda[y=0]}^{+} d\lambda \tag{22}$$

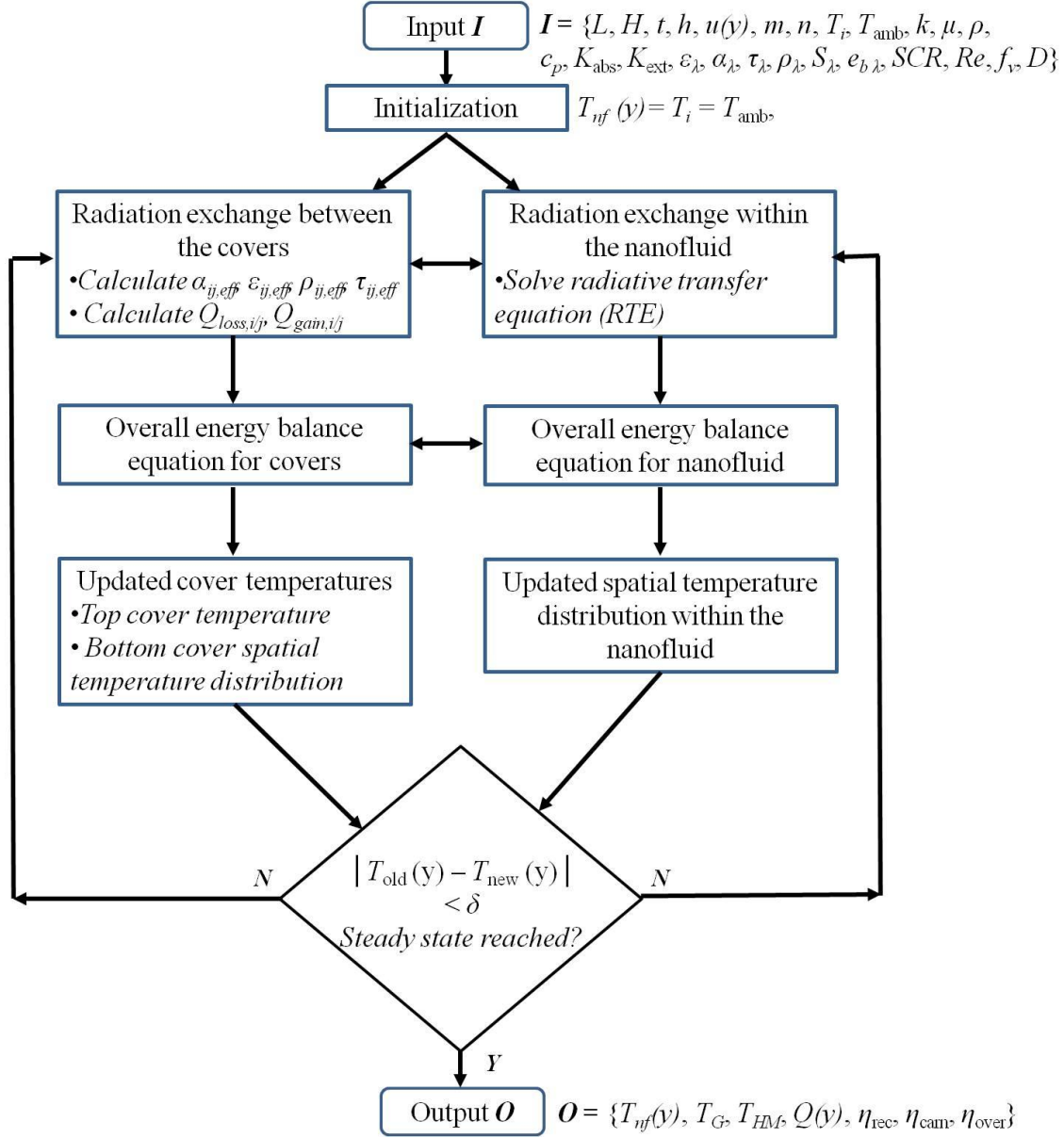

Fig. 8 Algorithm to calculate the performance parameters in relation to VARs.

***3.2.4 Surface absorption based solar thermal systems:*** Herein, the sunlight first interacts with the outer cover plate (plate 2, glass/heat mirror) and is subsequently absorbed by the plate 1 which happens to be a solar selective/black surface. This absorbed energy is then transferred to the fluid through conduction and convection. Radiation exchange between the two cover plates (one semi-transparent and one opaque in case of SAR) could be handled in a manner similar to that for VAR. The governing equation (and initial and boundary conditions) pertinent to the heat transfer processes in plate 1 and 2 are given by Eqs. (23) – (26) and Eq. (14) respectively.

$$\frac{\partial T}{\partial t} = \frac{k_1}{D_1 c_{p1}} \left[ \frac{\partial^2 T}{\partial x^2} + \frac{\partial^2 T}{\partial y^2} \right] + \frac{\int \tau_{2\lambda} \alpha_{1\lambda} S_\lambda d\lambda}{D_1 c_{p1}} \tag{23}$$

$$T(y)\big|_{t=0} = T_{amb} \tag{24}$$

$$Q\big|_{y=H+t_1} = \int \tau_{2\lambda} S_\lambda d\lambda - Q_{loss,1} \tag{25}$$

$$Q\big|_{y=H} = \frac{k_1}{D_1 c_{p1}} \left[ \frac{\partial T}{\partial y} \right] \tag{26}$$

The temperature distribution across the conduit depth and also along the conduit length is found by solving the overall energy equation [Eq. (27)] subject to initial and boundary conditions [Eqs. (28) – (30)].

$$\frac{\partial T}{\partial t} = \frac{k_f}{D_f c_{p,f}} \left[ \frac{\partial^2 T}{\partial y^2} \right] - u_y \frac{\partial T}{\partial x} \tag{27}$$

$$T(y)\big|_{t=0} = T_{amb} \tag{28}$$

$$Q\big|_{y=H} = \frac{k_f}{D_f c_{p,f}} \left[ \frac{\partial T}{\partial y} \right] \tag{29}$$

$$Q\big|_{y=0} = \frac{k_f}{D_f c_{p,f}} \left[ \frac{\partial T}{\partial y} \right] = 0 \tag{30}$$

All the aforementioned differential equations have been solved numerically using explicit form of finite difference technique. The details of the grid independence test and validation of the developed numerical models has been presented in appendices B and D respectively

## 4. Results and discussion

Present work delves into several aspects related to the working of VARs and SARs. In particular; the impact of enveloping surfaces, volume fraction of nanoparticles (in case of VARs), solar concentration ratios, Reynolds number, and inlet fluid temperatures on the performance characteristics of these receivers.

### 4.1 Performance characteristics of VARs and SARs

In order to clearly assess the performance characteristics in relation to scales of SCRs; we have categorized them into low (SCR ≤ 25) and medium-high (25 < SCR ≤ 100) SCR regimes. This shall help us to appreciate the fact that in each regime, there are different parameter combinations which influence the performance characteristics the most.

*4.1.1 Low solar concentration ratio regime (SCR ≤ 25):* In this regime we have analyzed the impact of nanoparticles volume fraction, receiver design, Reynolds number, and inlet fluid temperature on the performance characteristics of VARs and SARs

*Effect of nanoparticles volume fraction on efficiency of VARs:* Figure 9 shows surface plots for different designs of volumetric absorption-based receivers in low SCR regime. For all the three receiver designs, the receiver efficiency increases with volume fraction and attains a maximum value at $10^{-5}$ (in the present case); subsequently, the efficiency steeply decreases (or remains constant) with further increase in nanoparticles volume fraction. This happens because at very low volume fractions, radiation remains un-captured by the absorbing medium and at very high volume fractions, most of the radiation is captured near the surface (emulating surface absorption) and does not percolate down to the lower fluid layers in the receiver - resulting in low average fluid temperatures and increased thermal losses. Thus, there exists an optimum nanoparticle volume fraction that ensures favorable energy distribution across the fluid thickness. Moreover, in case of glass-heat mirror design, the heat mirror reflects the emitted radiation back to the glass (and nanofluid) - resulting in lower thermal losses. Further, it may be noted that receiver efficiency is higher at higher Reynolds number due to decrease in the emission losses. Effect of Reynolds number has been dealt with in a greater detail in the next section.

*Effect of Reynolds number on efficiency of VARs:* Figure 10 gives contour plots for different designs of volumetric absorption-based receivers at optimum volume fraction ($10^{-5}$). Optimum receiver efficiency values are achieved at higher Reynolds number and low solar concentration ratios. For any Reynolds number as the solar concentration ratio increases, the receiver efficiency decreases. However, there is very little variation in receiver efficiency for the case of glass-heat mirror design in this range of solar concentration ratios as is evident from the widely spaced contours. This is because at conditions of low Reynolds number and high concentration ratios, the temperatures at the top surface are high which exaggerate the emission losses. However, the design that includes a heat mirror reflects the emitted radiation back to the receiver thus reducing losses and therefore the receiver efficiency remains almost unaffected.

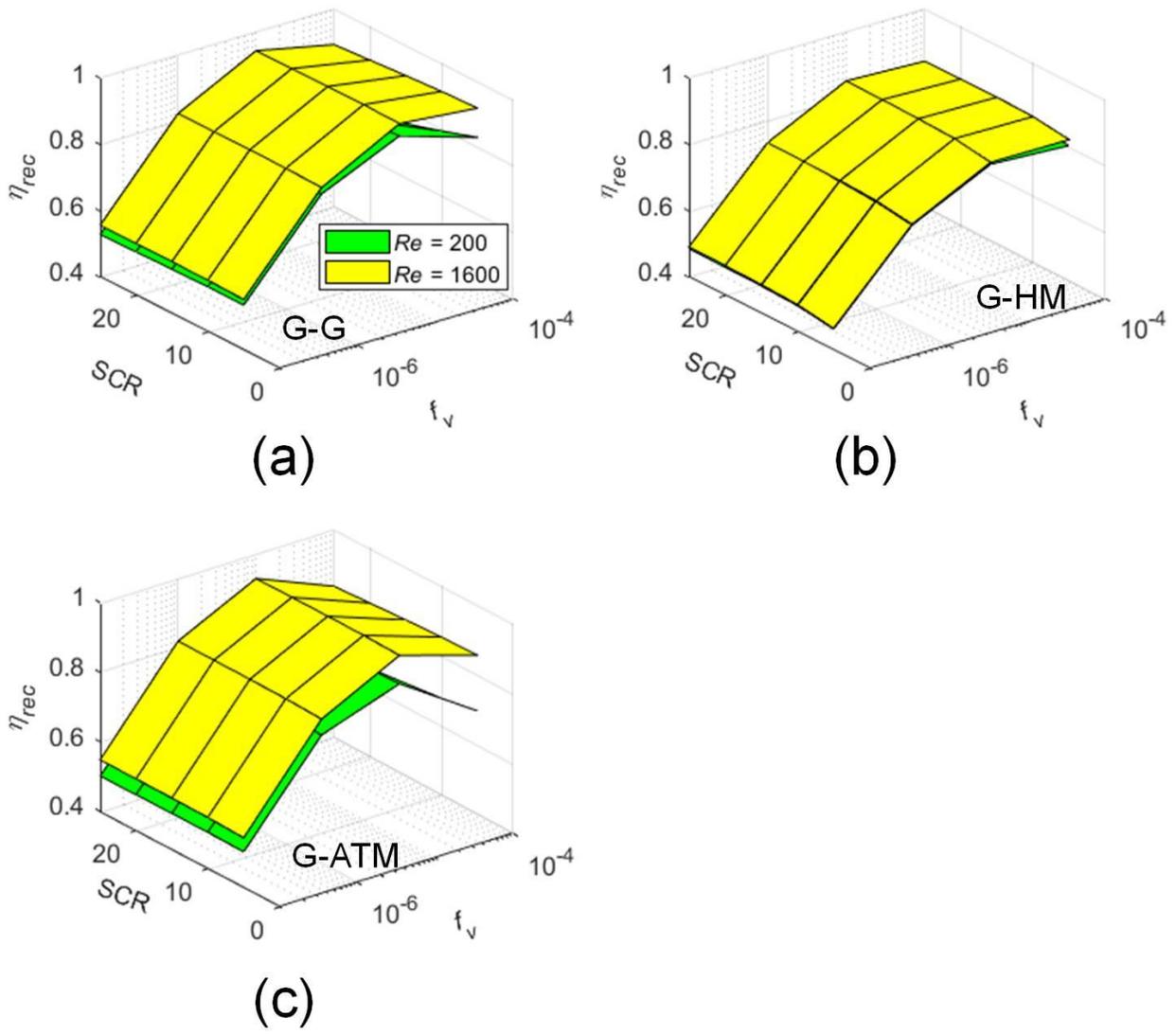

Fig. 9: Effect of nanoparticle volume fraction on receiver efficiency for volumetric absorption-based receivers in low SCR regime.

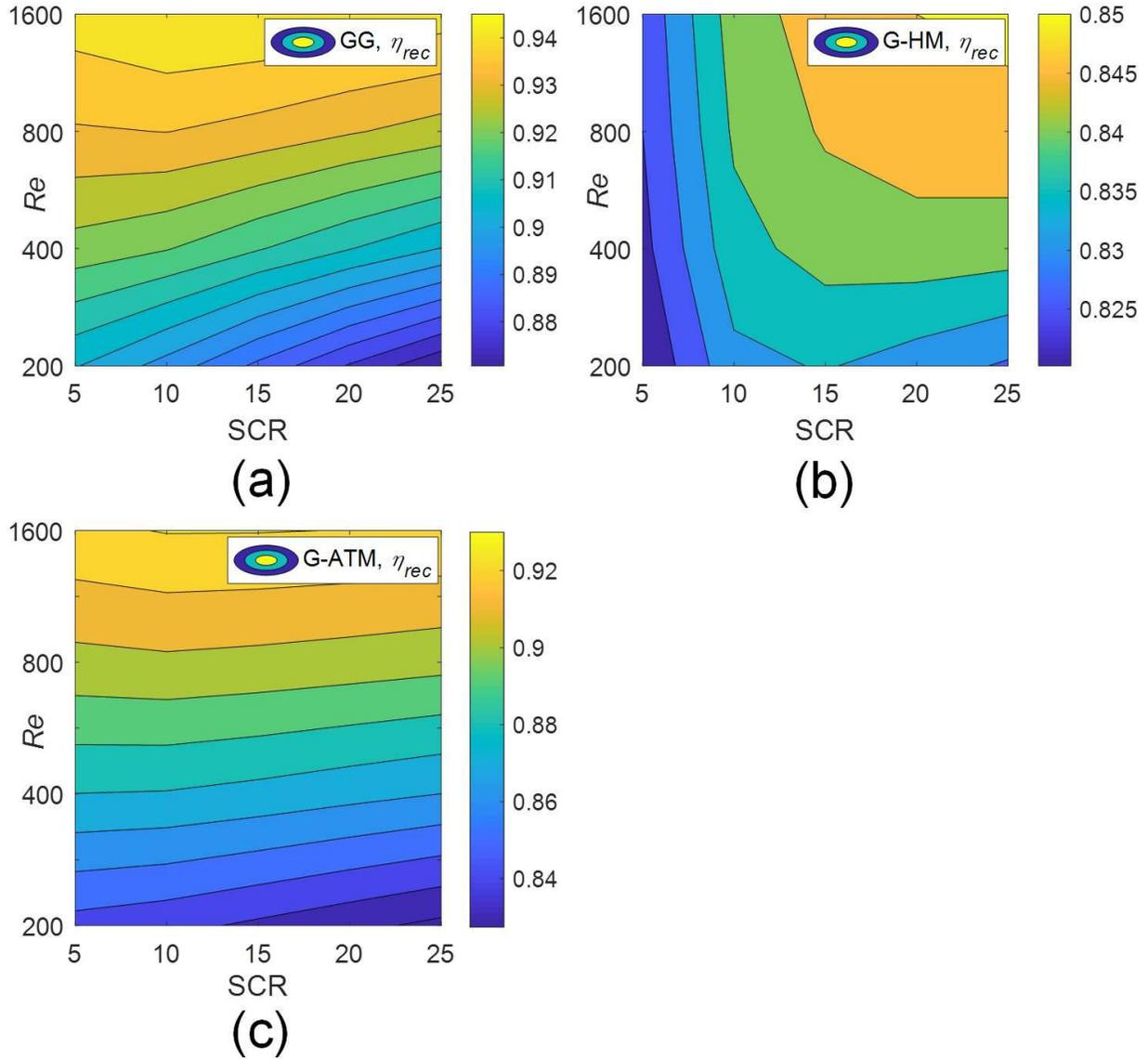

Fig. 10: Effect of Reynolds number on receiver efficiency of Volumetric absorption-based receiver in low SCR regime.

*Effect of inlet fluid temperature on receiver, Carnot and overall efficiency of VARs:*
Figure 11 compares the receiver efficiency, Carnot efficiency and overall efficiency for two designs of volumetric absorption-based receivers (i.e. the glass-heat mirror and the glass-glass design) at $Re$ = 200 and 1600. While the receiver efficiency is higher for glass-glass designs at low inlet fluid temperatures, it decreases at a more rapid rate as compared to that of glass-heat mirror designs at higher inlet temperatures. The Carnot efficiency curves for the two designs overlap with each other, so while the overall efficiency is higher for glass-glass design at low inlet fluid temperatures, at higher inlet fluid temperatures the overall efficiency is greater for glass-heat mirror design. The critical inlet fluid temperature after which the receiver efficiency for G-H exceeds the G-G depends on solar concentration ratio and it shifts to higher values of inlet fluid temperatures at higher SCR. This owes to the fact that at low inlet fluid temperatures the emission losses are not as much as they are at higher inlet temperatures and glass-heat mirror design reflects the emitted radiation back to the receiver.

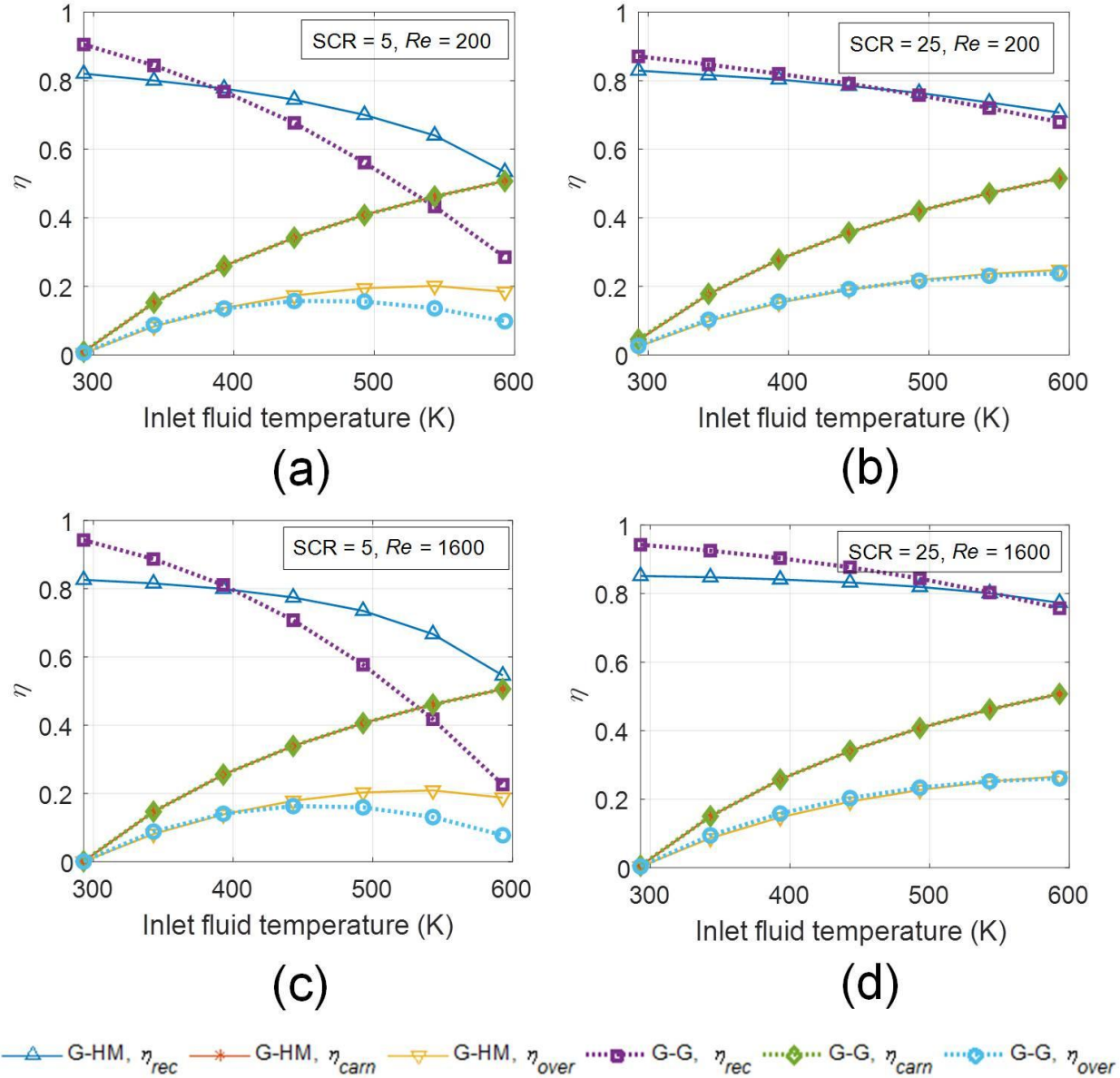

Fig. 11 Comparison of receiver, Carnot and overall efficiencies at different inlet fluid temperatures in case of VARs (a) SCR = 5, $Re$ = 200 (b) SCR = 25, $Re$ = 200 (c) SCR = 5, $Re$ = 1600, and (d) SCR = 25, $Re$ = 1600

*Effect of Reynolds number on efficiency of SARs:* Figure 12 gives contour plots for different designs of surface-absorption based receivers in low SCR regime. Just as was the case for volumetric-absorption receivers, optimum receiver efficiency values are attained at low solar concentration ratios and high Reynolds Number. However, the change in efficiency for the case of surface absorption-based receivers is much steeper than its volumetric absorption-based receiver counterparts as can be seen from the density of the contours which are much far apart in the latter case. This is because the top surface attains much higher temperature than the fluid owing to a thermal barrier created between the two thus making surface absorption-based receiver much more sensitive to changes in the solar concentration ratio.

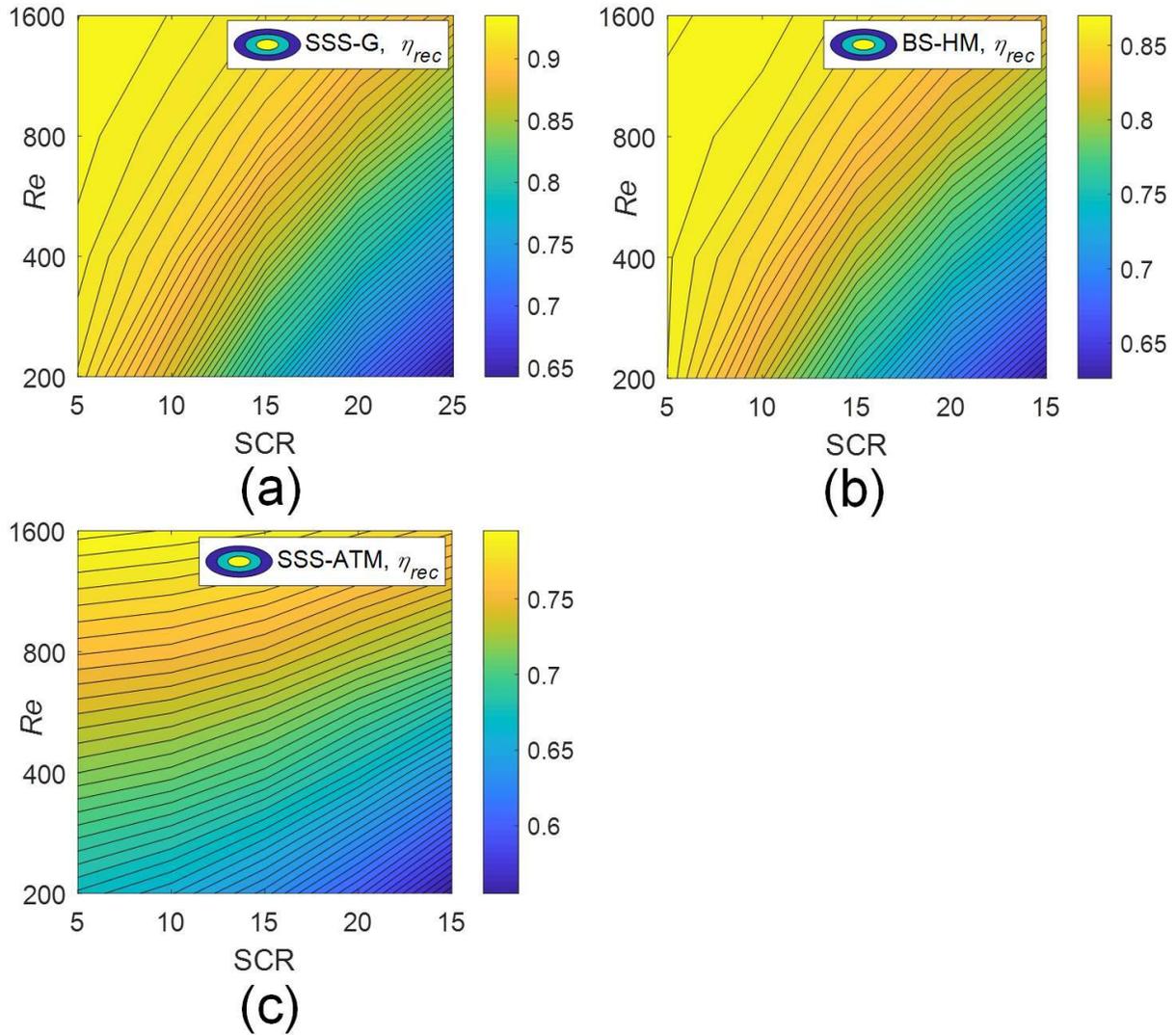

Fig. 12: Effect of Reynolds number on receiver efficiency of surface absorption-based receiver in low SCR regime.

*Effect of inlet fluid temperature on receiver, Carnot and overall efficiency of SARs:* Figure 13 compares the receiver efficiency, Carnot efficiency and overall efficiency for two designs of surface absorption-based receivers (i.e. the black surface-heat mirror and the solar selective surface-glass design) in low SCR regime at $Re$ = 200 and 1600. While the receiver efficiency is higher for solar selective surface-glass designs at low inlet temperatures, it decreases at a more rapid rate as compared to that of black surface-heat mirror designs at higher inlet temperatures. However at high solar concentration ratios, this difference is not very appreciable and receiver efficiencies remain almost the same for both the cases. The Carnot efficiency curves for the two designs overlap with each other. In the temperature range considered in the present study, the receiver efficiencies for black surface-heat mirror are always lower than those of solar selective surface-glass design. The receiver efficiencies are lower for the case of BS-HM design because of the lower amount of incident radiation that is allowed to pass through to the receiver. Also due to the presence of a solar selective surface in the SSS-G design, the radiation emitted by a solar selective surface is much lower than it is for a black surface in the BS-HM design.

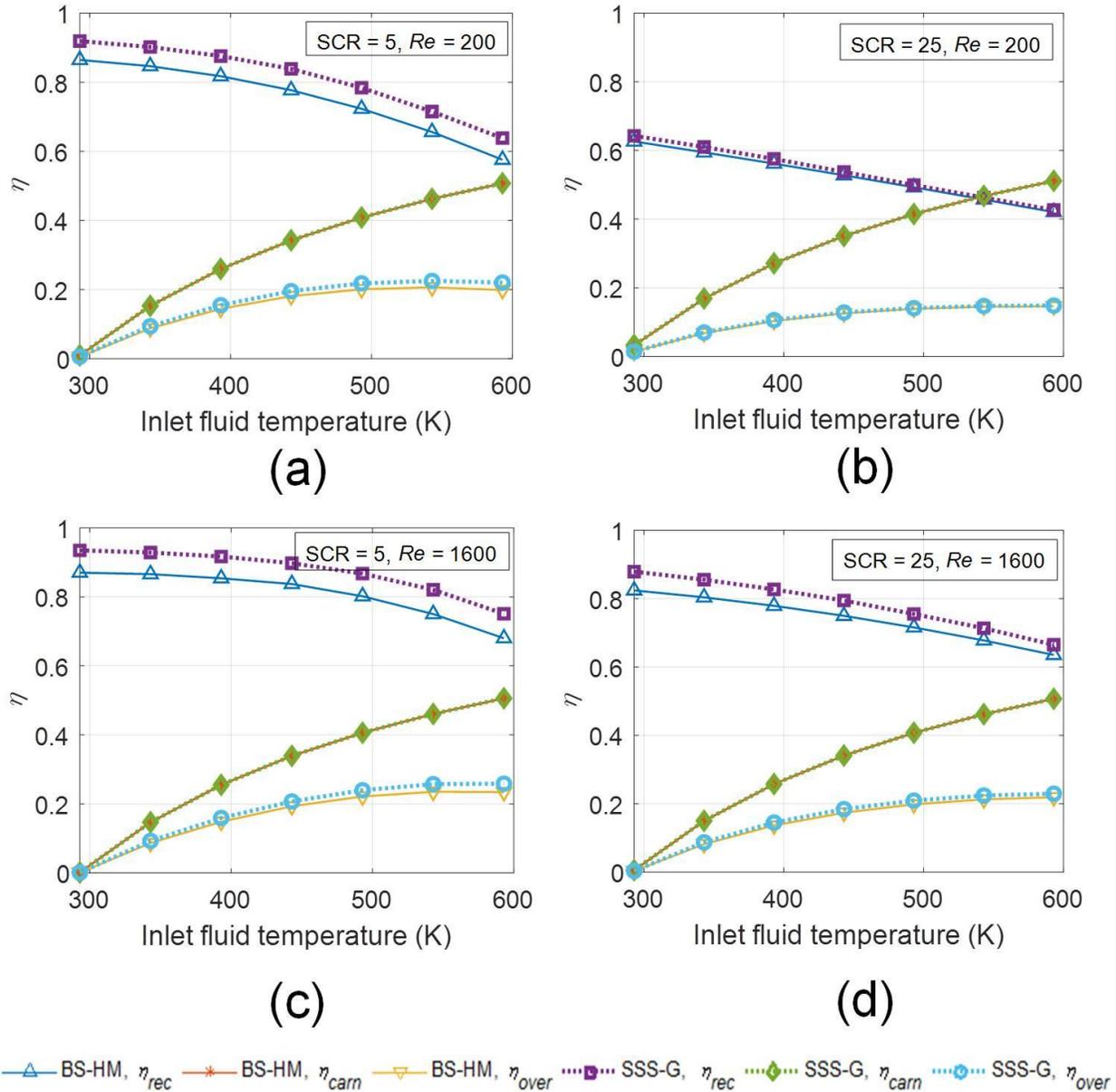

Fig. 13 Comparison of receiver, Carnot and overall efficiencies at different inlet fluid temperatures in case of SARs (a) SCR = 5, $Re$ = 200 (b) SCR = 25, $Re$ = 200 (c) SCR = 5, $Re$ = 1600, and (d) SCR = 25, $Re$ = 1600

***4.1.2 Medium-high solar concentration ratio regime (25 < SCR ≤ 100):*** In this regime, we have analyzed the impact of inlet fluid temperature, receiver design, and Reynolds number on the performance characteristics of VARs and SARs.

*Effect of inlet fluid temperature and Reynolds number on receiver, Carnot and overall efficiency of VARs:* Figure 14 compares the receiver efficiency, Carnot efficiency and overall efficiency for two designs of volumetric absorption-based receivers (i.e. the glass-heat mirror and the glass-glass design) at *Re* = 200 and 1600. The receiver efficiencies do not vary much (although slightly decrease in all the cases) with increase in inlet fluid temperatures. This may be attributed to the fact that at high SCRs, the relative magnitude of useful thermal energy gain outpowers the thermal energy losses – thus limiting the efficiency decrease with increase in inlet fluid temperatures. However, similar to the trennds in low SCR regime; the Carnot efficiency curves for the two designs overlap with each other.

Further, while the overall efficiency is higher for glass-glass design across the entire inlet fluid temperatures range; at still higher inlet fluid temperatures the overall efficiency is greater for glass-heat mirror design. Moreover, the critical inlet temperature where the receiver efficiency for G-H starts to overtake the G-G case shifts to much higher inlet fluid temperatures (~ 600K). Clearly pointing out that the critical point shifts towards right with increase in SCR.

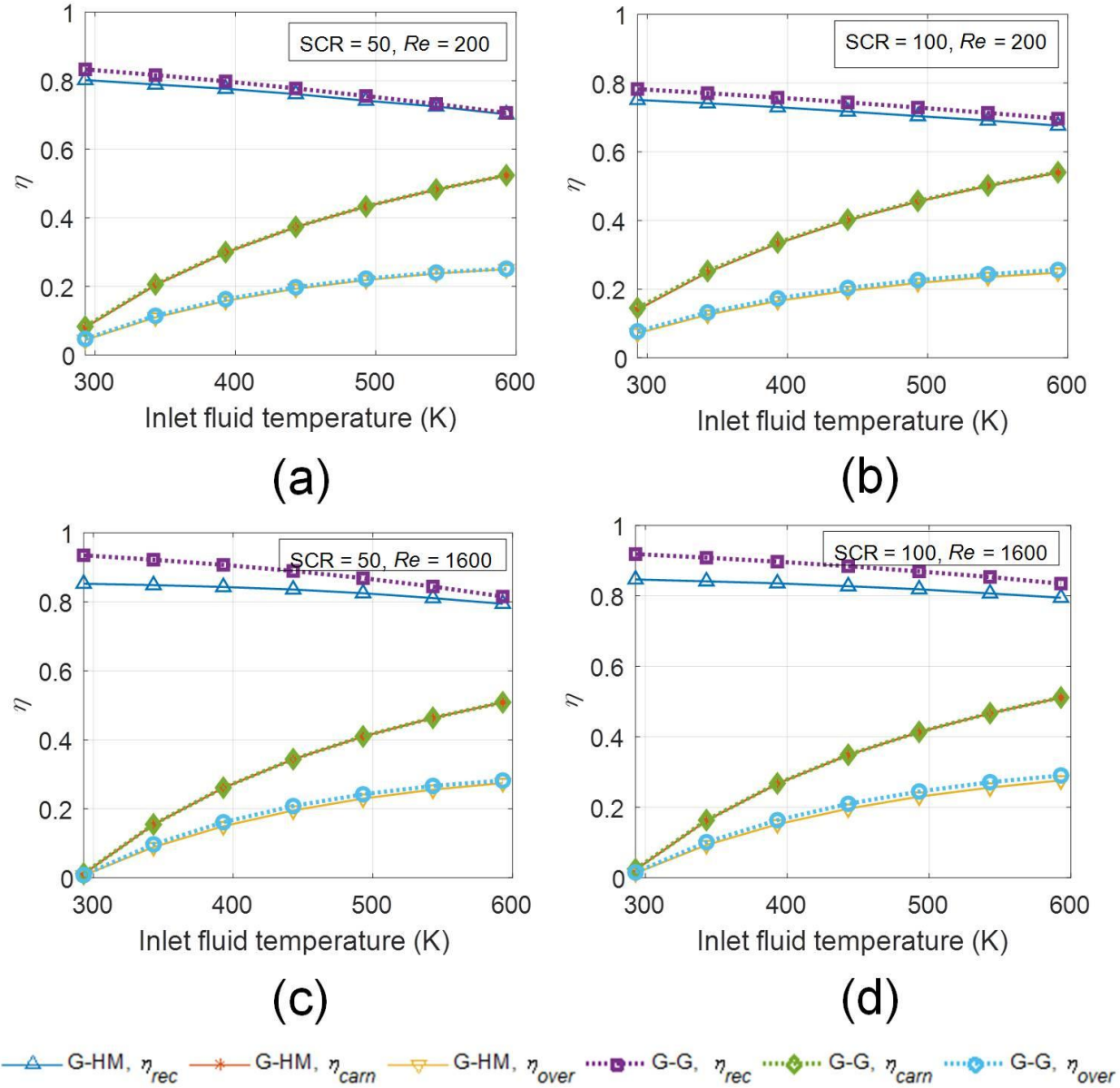

Fig. 14 Comparison of receiver, Carnot and overall efficiencies at different inlet fluid temperatures in case of VARs (a) SCR = 50, $Re$ = 200 (b) SCR = 100, $Re$ = 200 (c) SCR = 50, $Re$ = 1600, and (d) SCR = 100, $Re$ = 1600

*Effect of inlet fluid temperature and Reynolds number on receiver, Carnot and overall Efficiency of SARs:* Figure 15 compares the receiver efficiency, Carnot efficiency and overall efficiency for two designs of surface absorption-based receivers (i.e. the black surface-heat mirror and the solar selective surface-glass design) at $Re$ = 200 and 1600. At low Reynolds number ($Re$ = 200), the efficiecies are considerably low even at low inlet fluid temperatures. Moreover, efficiencies further decrease with increase in inlet fluid temperatures and SCR. This can attributed to the fact that at low Reynolds numbers and high SCRs, staggering amount of thermal energy gets accumulated at the surface – owing to low Reynolds number, the fluid is unable to take this energy with it, therefore resulting in huge temperature overheat (i.e., surface temperatures are significantly higher than the fluid temperatures). This further leads to escalation of thermal losses. Simillar trends could be seen at high Reynolds number as well; however, the magnitude of temperature overheat and subsequent emission losses being dimnished in comparison to the low Reynolds number case.

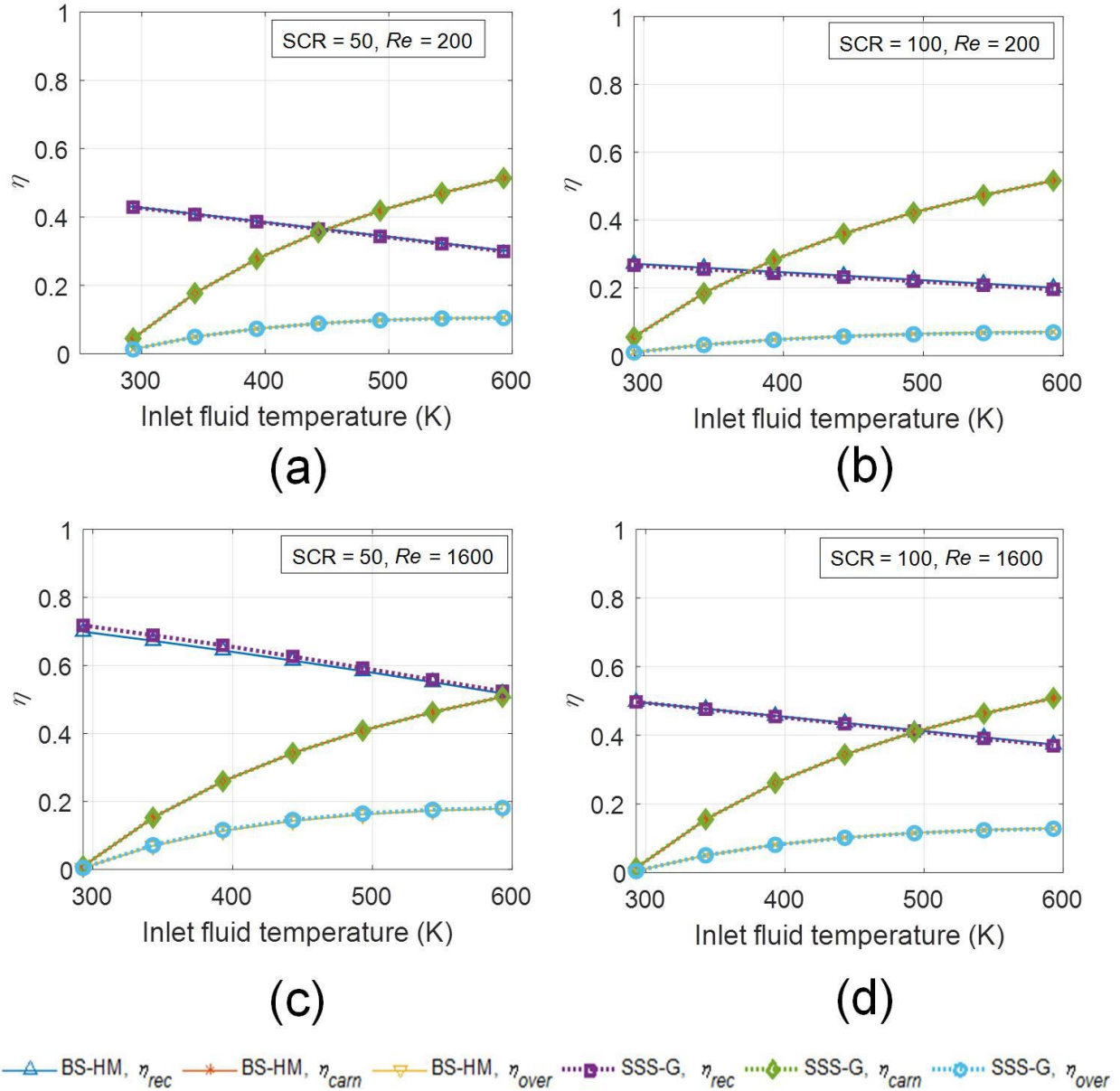

Fig. 15 Comparison of receiver, Carnot and overall efficiencies at different inlet fluid temperatures in case of SARs (a) SCR = 50, $Re$ = 200 (b) SCR = 100, $Re$ = 200 (c) SCR = 50, $Re$ = 1600, and (d) SCR = 100, $Re$ = 1600

**4.2 Delineating optimal range of operating and design parameters for VARs and SARs: The big picture**

Figure 16 shows how receiver, Carnot and overall efficiencies compare between volumetric and surface absorption based receivers. It is seen that as the solar concentration ratios increase, the receiver efficiencies drop rapidly for surface absorption-based receivers (both for $T_{in}$ = 293K and 593K). Whereas, in case of VARs, we have two distinct trends: while efficiencies decrease slightly or remain almost constant) at low inlet fluid temperatures ($T_{in}$ = 293K); at high inlet fluid temperatures ($T_{in}$ = 593K), efficiencies first increase and then stagnate with increase in SCR.

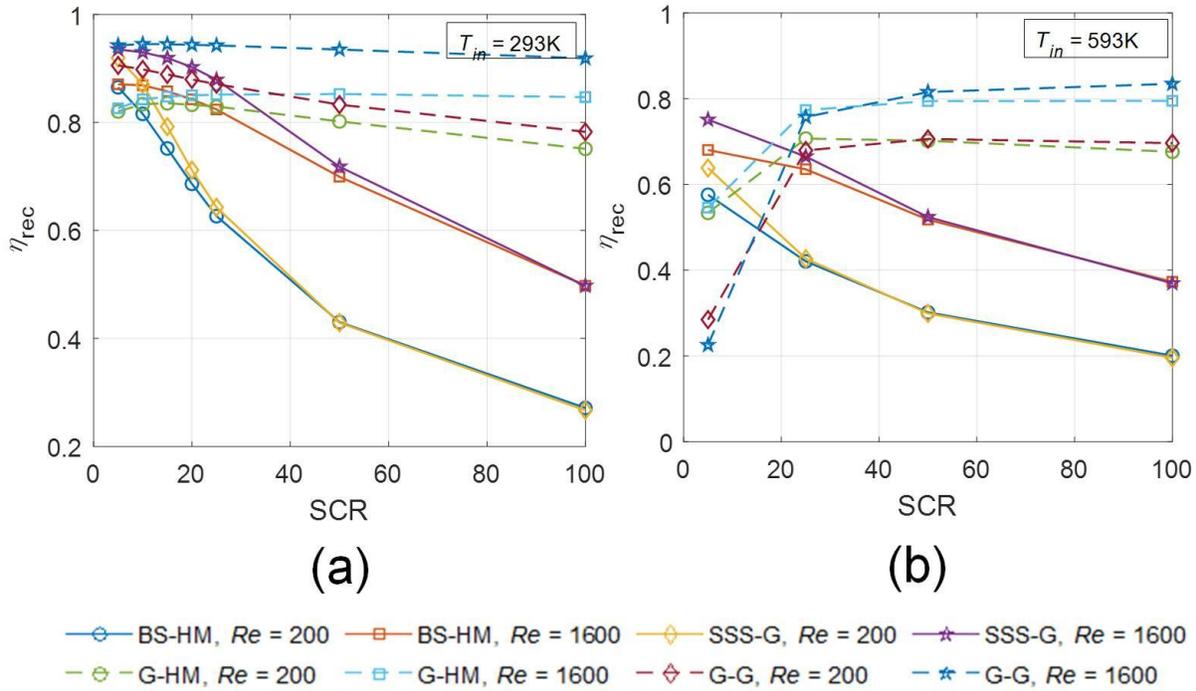

Fig. 16 Difference in efficiencies between volumetric and surface absorption based receivers (a) receiver efficiency, $T_{in}$ = 293K, and (b) receiver efficiency, $T_{in}$ = 593K.

In order to clearly understand the heat transfer mechanisms involved in VARs and SARs, we have plotted the exit temperature field for these at different Reynolds number and inlet fluid temperatures. Figure 17 compares the exit temperature field plots for black surface-heat mirror design (surface absorption-based receiver) with those of glass-heat mirror design (volumetric absorption-based receiver). Particularly, G-HM and BS-HM cases have been considered as these are similar in all aspects except for the mechanisms of heat transfer involved. One can see that the maximum temperature for the case of SAR design is at the surface and there is a very steep difference in temperature between the surface and the topmost layer of the fluid (i.e. significant overheat temperature exists). On the contrary, the temperature of the top glass plate for the case of VAR design is lower than the temperature of topmost layer of fluid in the conduit (i.e., there exists a temperature inversion in case of VARs) which leads to lower emission losses and better distribution of energy across the fluid layers – resulting in higher efficiencies for volumetric absorption-based receivers.

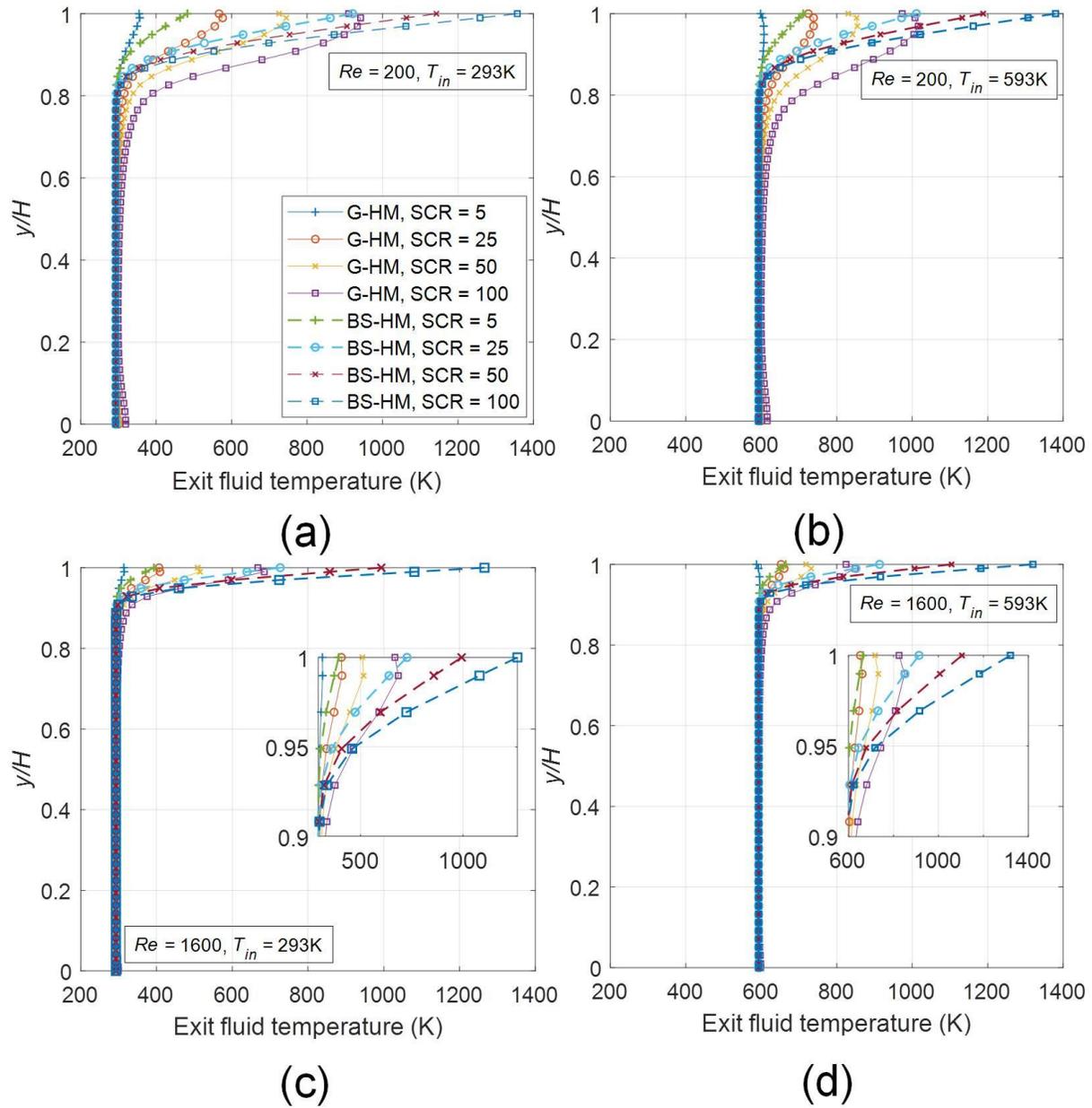

Fig. 17: Comparison of exit temperature distribution between volumetric and surface absorption based receivers (a) $Re = 200$, $T_{in} = 293K$ (b) $Re = 200$, $T_{in} = 593K$ (c) $Re = 1600$, $T_{in} = 293K$, and (d) $Re = 200$, $T_{in} = 593K$.

Therefore, the absence of temperature overheat and instead existance of temperature inversion in case of VARs lends them to be more suitable (particularly in medium-high SCRs regime) as compared to corresponding SARs. Receiver efficiency enhancements on the order of ~35% @ $Re = 1600$ to 49% @ $Re = 200$ could be achieved in case of VARs relative to their SAR counterparts at SCR = 100.

Table 1 seves as a general guideline for selection of solar thermal systems in various operating regimes. Clearly, depending on the operating conditions, appropriate VAR/SAR design variant should be carefully chosen to ensure high receiver efficiencies.

Table 1 Receiver efficiencies for VAR/SAR design variants in various operating regimes.

| @ $T_{in}$ = 293K | $Re$ = 200 | $Re$ = 1600 |
|---|---|---|
| SCR ≤ 25 | $\eta_{G-G} > \eta_{G-HM} > \eta_{SSS-G} > \eta_{BS-HM}$ | $\eta_{G-G} > \eta_{SSS-G} > \eta_{BS-HM} > \eta_{G-HM}$ |
| 25 < SCR ≤ 100 | $\eta_{G-G} > \eta_{G-HM} > \eta_{SSS-G} > \eta_{BS-HM}$ | $\eta_{G-G} > \eta_{G-HM} > \eta_{SSS-G} > \eta_{BS-HM}$ |
| @ $T_{in}$ = 593K | $Re$ = 200 | $Re$ = 1600 |
| SCR ≤ 15 | $\eta_{G-HM} > \eta_{SSS-G} > \eta_{BS-HM} > \eta_{G-G}$ | $\eta_{SSS-G} > \eta_{BS-HM} > \eta_{G-HM} > \eta_{G-G}$ |
| 15 < SCR ≤ 25 | $\eta_{G-HM} > \eta_{G-G} > \eta_{SSS-G} > \eta_{BS-HM}$ | $\eta_{G-HM} > \eta_{G-G} > \eta_{SSS-G} > \eta_{BS-HM}$ |
| 25 < SCR ≤ 100 | $\eta_{G-G} \approx \eta_{G-HM} > \eta_{SSS-G} \approx \eta_{BS-HM}$ | $\eta_{G-G} > \eta_{G-HM} > \eta_{SSS-G} > \eta_{BS-HM}$ |

## 5. Conclusions and future work

On the whole, the present work presents a comprehensive theoretical modeling framework that is generic enough to include intricate coupled heat transfer phenomena for both volumetric as well as surface absorption based solar thermal systems. We have been able to provide detailed guidelines for selecting a particular receiver design variant under given set of operating conditions in laminar flow regime. This shall serve as a refernce document for designing and further improving upon the performance of solar therrmal systems. As a part of future work, it is envisaged that a comprehensive modeling framework for turbulent flow regime is also developed.

## Acknowledgements

This work is supported by DST-SERB (Sanction order no. ECR/2016/000462). AS and VK also acknowledge the support provided by Mechanical Engineering Department at TIET Patiala. MK acknowledges the support provided by Mechanical Engineering Department at MNIT Jaipur.

## Appendix
### A: Mathematical modeling of spectral optical properties
*Glass*

The transmissivity values for low-iron glass have been calculated utilizing data (optical constants '$n$' and '$\kappa$') from Ref. [41]. Once the values of '$n$' and '$\kappa$' are known, the reflectivity, transmissivity and absorptivity values for glass can be evaluated through the following steps:
The spectral internal transmissivity '$\tau_a$' can be calculated using,

$$\tau_{a,\lambda} = e^{-K_{e,\lambda} ds}, \tag{A1}$$

where,

$$K_{e,\lambda} = \frac{4\pi\kappa}{\lambda}, \tag{A2}$$

The effective spectral reflectance is evaluated from the Fresnel relations:

$$\rho_{\perp,\lambda} = \frac{(n_{vac}\cos\psi - w)^2 + v^2}{(n_{vac}\cos\psi + w)^2 + v^2}$$

$$\rho_{\parallel,\lambda} = \frac{\left[(n_{h,\lambda}^2 - \kappa_{h,\lambda}^2)\cos\psi - n_{vac}w\right]^2 + \left[2n_{h,\lambda}\kappa_{h,\lambda}\cos\psi - n_{vac}v\right]^2}{\left[(n_{h,\lambda}^2 - \kappa_{h,\lambda}^2)\cos\psi + n_{vac}w\right]^2 + \left[2n_{h,\lambda}\kappa_{h,\lambda}\cos\psi + n_{vac}v\right]^2}, \quad (A3)$$

$$\rho_{avg,\lambda} = \frac{\rho_{\perp,\lambda} + \rho_{\parallel,\lambda}}{2}$$

where,

$$2w^2 = \left(n_{h,\lambda}^2 - \kappa_{h,\lambda}^2 - n_{vac}^2\sin^2\psi\right) + \sqrt{\left(n_{h,\lambda}^2 - \kappa_{h,\lambda}^2 - n_{vac}^2\sin^2\psi\right)^2 + 4n_{h,\lambda}^2\kappa_{h,\lambda}^2}$$

$$2v^2 = -\left(n_{h,\lambda}^2 - \kappa_{h,\lambda}^2 - n_{vac}^2\sin^2\psi\right) + \sqrt{\left(n_{h,\lambda}^2 - \kappa_{h,\lambda}^2 - n_{vac}^2\sin^2\psi\right)^2 + 4n_{h,\lambda}^2\kappa_{h,\lambda}^2} \quad (A4)$$

The spectral absorptivity for glass is then calculated from

$$a_\lambda = \varepsilon_\lambda = 1 - \tau_{a,\lambda} - \rho_{avg,\lambda}, \quad (A5)$$

The values calculated were scaled such that the solar weighted transmissivity of glass is 97.8%.

*Nanofluid*

As radiation propagates through the nanofluid; its magnitude changes owing to absorption and scattering mechanisms. Attenuation of radiation due to scattering is negligibly small as the particles are much smaller than the wavelength of incident light which means that scattering occurs in the Rayleigh regime and absorption dominates [29]. Absorption and scattering mechanisms could be quantified using the following expressions:

$$K_{a,\lambda,bf} = \frac{4\pi\kappa}{\lambda}, \quad (A6)$$

where '$K_{a,\lambda}$' is the spectral absorption coefficient, '$\kappa$' is the spectral index of absorption, and '$\lambda$' is the wavelength.

The combined attenuation from absorption and scattering by nanoparticles is taken care of by extinction coefficient, given as,

$$K_{e,\lambda,np} = \frac{1.5 f_v Q_{e,\lambda}(\beta,m)}{d}, \quad (A7)$$

where, '$K_{e,\lambda}$' is the spectral extinction coefficient, '$f_v$' is the volume fraction, '$d$' is the characteristic dimension (hydrodynamic diameter, $d = 1$nm), '$\beta$' is the size parameter, '$m$' is the normalized refractive index and '$Q_{e,\lambda}$' is the extinction efficiency of the particles defined as follows [22]:

$$\beta = \frac{\pi d}{\lambda}, \quad (A8)$$

$$m = \frac{n_{np}}{n_{bf}}, \tag{A9}$$

$$Q_{e,\lambda} = 4\beta \operatorname{Im}\left\{\frac{m^2-1}{m^2+2}\left[1+\frac{\beta^2}{15}\left(\frac{m^2-1}{m^2+2}\right)\frac{m^4+27m^2+38}{2m^2+3}\right]\right\}+\frac{8}{3}\beta^4 \operatorname{Re}\left\{\left(\frac{m^2-1}{m^2+2}\right)^2\right\}, \tag{A10}$$

The values of '$\kappa$' for base-fluid were obtained from Ref. [43]. The '$n$' and '$\kappa$' values for nanoparticles have been taken from Ref. [44, 45].

Once we have calculated the value of extinction coefficients for base-fluid and nanoparticles individually, we can find the value of extinction coefficient for the nanofluid simply by adding the two.

$$K_{e,\lambda,nf} = K_{a,\lambda,bf} + K_{e,\lambda,np}, \tag{A11}$$

**B: Numerical modeling**

Plate 1 (solar selective/black/glass) and the fluid (basefluid/nanofluid) have been discretized into finite control volumes. Explicit form of finite difference technique has been employed to convert the integro-differential equations into set of algebraic equations – characteristic algebraic equations for interior and boundary nodes. Figure B1 show the discretization strategy followed for numerical modeling VARs in particular. RTE is essentially solved in $y$-$z$ plane; whereas, the overall energy balance equation is solved in $x$-$y$ plane until steady state is reached.

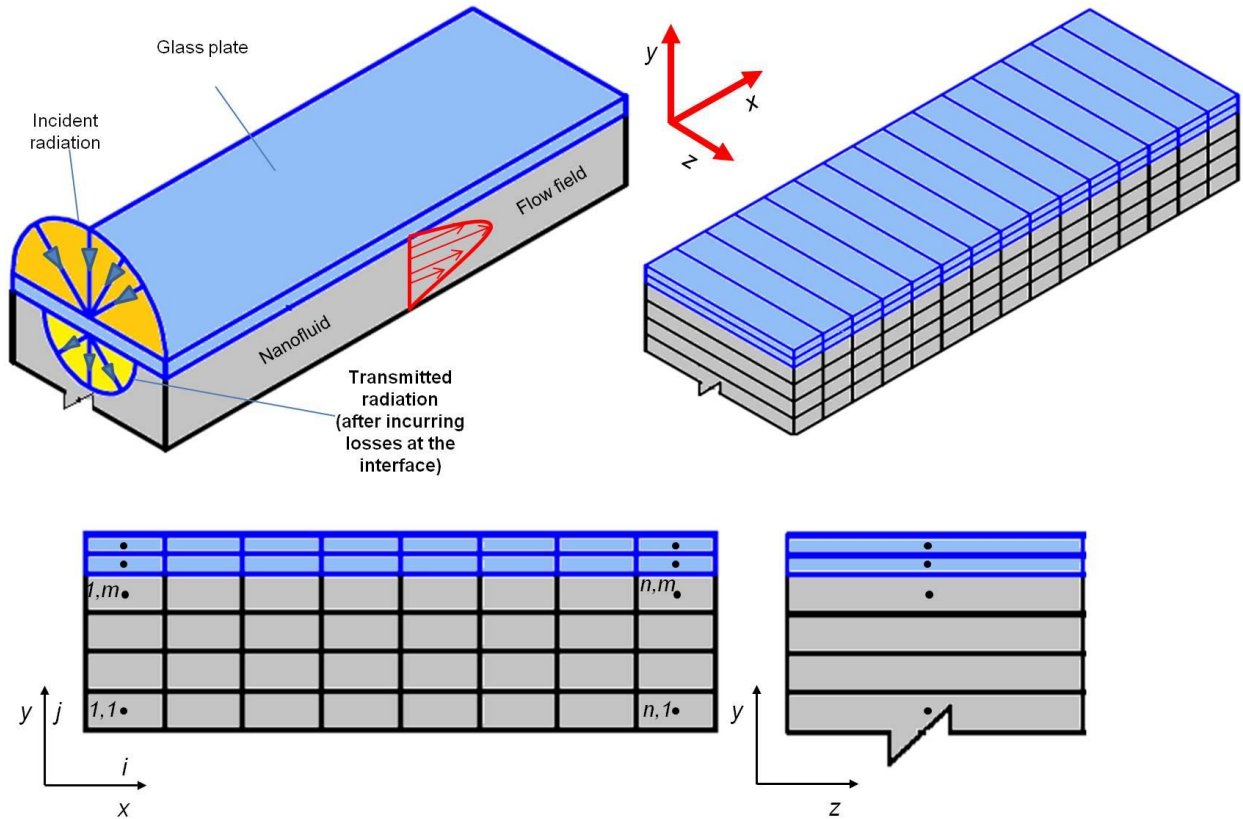

Fig. B1 Discretization strategy for numerical modeling of VARs.

Further, It may be noted that grid independence test was carried out and it was found that the results were nearly independent of grid size beyond 49×700 (see Fig. B2).

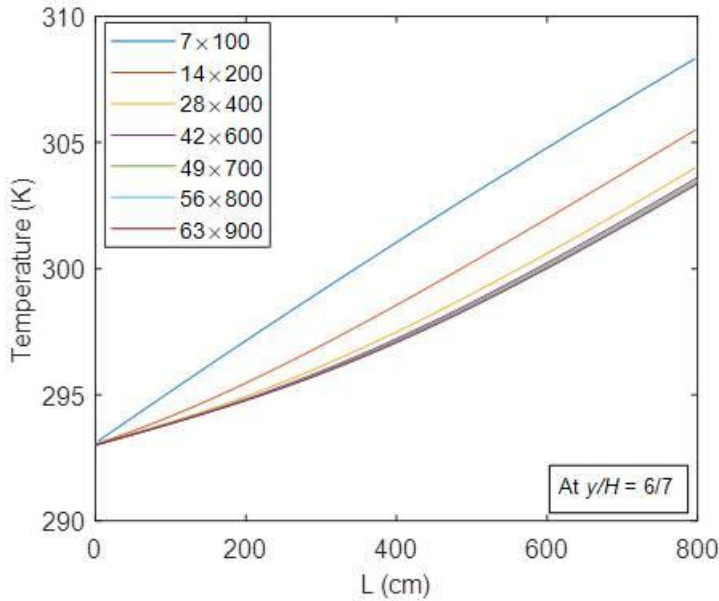

Fig. B2 Spatial temperature distribution along the conduit length at $y/H = 6/7$ for various grid sizes.

**C: Solution of the radiative transfer equation (RTE)**

Figure C1 shows the *y-z* plane of the conduit. The nanofluid is bounded by glass and reflective plates at the top and bottom respectively. To find the values of radiation intensity in positive and negative directions at various points along the depth of the channel the RTE has been solved numerically. Once intensity at different depths in different directions (+ve and -ve) is known, the 'divergence' which is the net radiative energy per unit time and volume leaving a differential control volume into which the channel depth has been subdivided can be found.

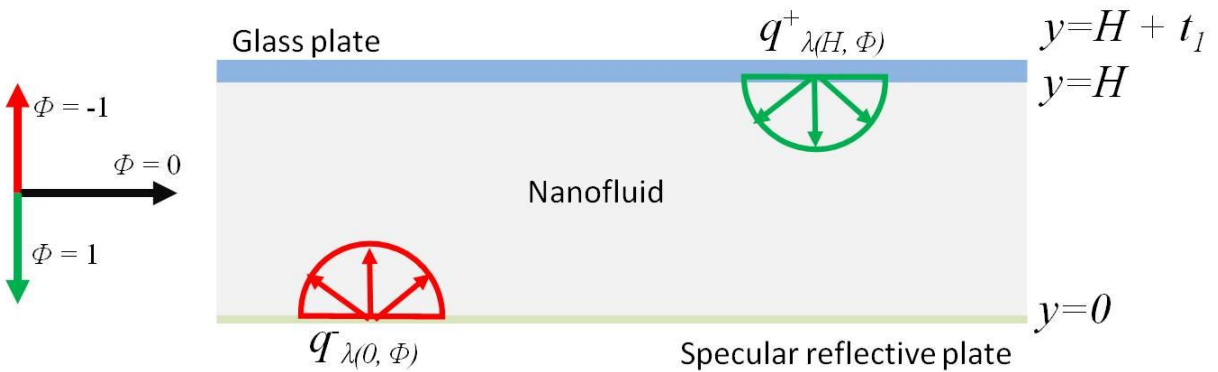

Fig. C1 Schematic showing the y-z plane defining the +ve and -ve directions of $\Phi$ and the heat flux,

From the boundary conditions we get Eqs. (C1) and (C2).

$$I^{+}_{\lambda(H,\phi)} = S_{\lambda 1}.\tau_{\lambda 1} + \rho_{\lambda 1}.I^{-}_{\lambda(H,-\phi)} + \alpha_{\lambda 1}.I_{b\lambda[y=H]} \tag{C1}$$

$$I^{-}_{\lambda(0,\phi)} = \rho_{\lambda 3}.I^{+}_{\lambda(0,+\phi)} \tag{C2}$$

We get two more equations from the equation of radiative transfer for an absorbing, emitting non-scattering medium [Eqs. (C3) and (C4)]

$$I^{-}_{\lambda(H,\phi)} = I^{-}_{\lambda(0,\phi)} \exp\left[\frac{\theta_H}{\phi}\right] - \int_0^{\theta_H} \frac{I_{b(\theta')}}{\phi} e^{\left(\frac{\theta_H}{\phi}\right)} d\theta' \tag{C3}$$

$$I^{+}_{\lambda(0,\phi)} = I^{+}_{\lambda(H,\phi)} \exp\left[-\frac{\theta_H}{\phi}\right] + \int_{\theta_H}^{0} \frac{I_{b(\theta')}}{\phi} e^{\left(\frac{\theta'-\theta_H}{\phi}\right)} d\theta' \tag{C4}$$

Radiation travelling from the top cover to the bottom is regarded as +ve and vice versa.
Solving Eqs. (C1) - (C4), we get Eqs. C5 and C6

$$I^{+}_{\lambda(H,\phi)} = \frac{S_{\lambda 1}.\tau_{\lambda 1} + \alpha_{\lambda 1}.I_{b\lambda T_{[y=H]}} + \rho_{\lambda 1}.\rho_{\lambda 3}.\exp\left[\frac{-\theta_H}{\phi}\right].\int_{\theta_H}^{0} \frac{I_{b\lambda(\theta')}}{\phi}.e^{\left(\frac{-\theta'}{\phi}\right)} d\theta' + \rho_{\lambda 1}.\int_{0}^{\theta_H} \frac{I_{b\lambda(\theta')}}{\phi}.e^{\left(\frac{\theta_H-\theta'}{\phi}\right)} d\theta'}{\left(1 - \rho_{\lambda 1}.\rho_{\lambda 3}.\exp\left[\frac{-2\theta_H}{\phi}\right]\right)} \tag{C5}$$

for $\phi > 0$

$$I^{-}_{\lambda(0,\mu)} = \frac{\rho_{\lambda 3}.S_{\lambda 1}.\tau_{\lambda 1}.\exp\left[\frac{\theta_H}{\mu}\right] + \alpha_{\lambda 1}.\rho_{\lambda 3}.I_{b\lambda[y=H]}.\exp\left[\frac{\theta_H}{\mu}\right] - \rho_{\lambda 1}.\rho_{\lambda 3}.\exp\left[\frac{\theta_H}{\mu}\right].\int_{\theta_H}^{0} \frac{I_{b\lambda(\theta')}}{\mu}.e^{\left(\frac{\theta'}{\mu}\right)} d\theta' - \rho_{\lambda 3}.\int_{0}^{\theta_H} \frac{I_{b\lambda(\theta')}}{\mu}.e^{\left(\frac{\theta_H-\theta'}{\mu}\right)} d\theta'}{\left(1 - \rho_{\lambda 1}.\rho_{\lambda 3}.\exp\left[\frac{2\theta_H}{\mu}\right]\right)} \tag{C6}$$

for $\phi < 0$

The hemispherical flux at any location is obtained by integrating the intensity field.

$$q^{+}_{\lambda} = 2\pi \int_{0}^{1} I^{+}_{\lambda} \phi d\phi$$
$$q^{-}_{\lambda} = 2\pi \int_{-1}^{0} I^{-}_{\lambda} \phi d\phi \tag{C7}$$

Once the values of intensity at the boundaries are calculated, one can use Eq. (C8) to find the intensity and thus hemispherical flux at any point in between the two plates.

$$I^{+}_{\lambda}(\theta) = I^{+}_{1\lambda} \exp\left(\frac{-\theta}{\phi}\right) + \int_{0}^{t} I_{b\lambda}(\theta').\exp\left(\frac{\theta'-\theta}{\phi}\right) \frac{d\theta'}{\phi} \tag{C8}$$

The spectral intensity of radiation emitted locally inside a medium is given by Eq. (C9)

$$I_{b,\lambda} = \frac{C_1 n^3}{\lambda^5 \left(e^{C_2/\lambda T} - 1\right)} \tag{C9}$$

where, $C_1 = 3.74177 \times 10^8$ and $C_2 = 3.74177 \times 10^8$, '$n$' is the refractive index, '$\lambda$' is wavelength and '$T$' is for temperature (Howell).

The divergence for any generic volume element is given by Eq. (C10) as

$$Q_{rad,nf} = q_{net,out} - q_{net,in} \tag{C10}$$

The divergence thus calculated can be included in the energy equation to find the temperature distribution within the nanofluid.

**D: Validation of numerical models**

RTE model validation
Two infinite black walls at a distance '$H$' are considered. The known conditions are the temperatures ($T_1$ and $T_2$) and emissivities ($\varepsilon_1 = \varepsilon_2 = 1$) of the two walls, and the absorption coefficients ($K_a$) of the intervening medium. The non-dimensional blackbody flux at any optical thickness '$t$' (where $t = K_a y$, $y$ being the distance from wall 1) is given by Eq. (D1) as

$$e_b^*(t) = \frac{e_b(t) - e_{b2}}{e_{b1} - e_{b2}} \tag{D1}$$

The value of $e_b^*$ at the temperature of wall 1 is $e_b^*(T_1) = 1$ and the value of $e_b^*$ at the temperature of wall 2 is $e_b^*(T_2) = 0$. Figure D1 shows the comparison between present developed model and that in Ref. [46]; clearly the two closely match for various values of absorption coefficients.

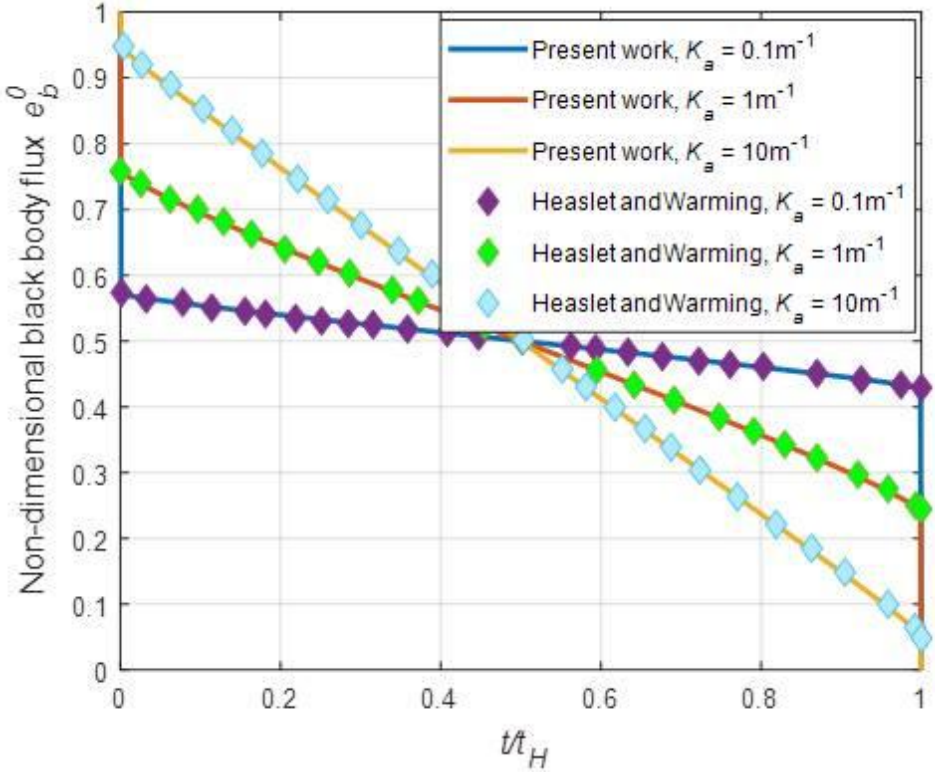

Fig. D1 Non-dimensional temperature profile for a gray slab in radiative equilibrium for different values of optical thickness - comparison of results of present work with those in Ref. [46].

Validation of numerical model for SARs

In order to validate the numerical model developed for SARs, the results obtained from the numerical model were compared with those obtained from ANSYS®. A channel length of 4 meters and width of 5cm has been considered with a black surface at the top. Solar radiation (1000Wm$^{-2}$) falls on the top surface of the channel absorbing all radiation and heating up, thus in turn heating the fluid flowing through the channel. The initial temperature of the fluid and the ambient temperatures are both assumed to be 293K. Temperature distribution was obtained for both cases. Figure D2 shows a comparison of temperature distribution of both cases along the channel length at various channel heights. Both the results are in agreement with each other thus validating the numerical approach for finding the temperature distribution in a surface absorption-based receiver.

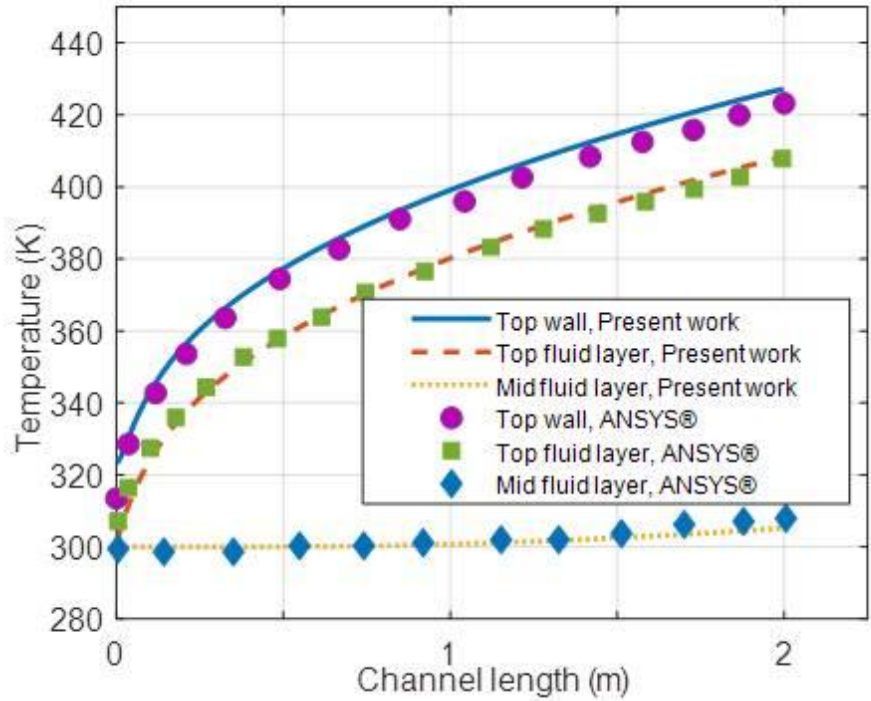

Fig. D2 Comparison of temperature distribution at different channel heights as obtained from the developed numerical model and those obtained from *ANSYS® Academic Research Mechanical, Release 18.1*.

## Nomenclature

English Symbols:
- $c_p$   specific heat [J kg$^{-1}$ K$^{-1}$]
- D    density [kg/m$^3$]
- $d$    characteristic dimension for nanoparticles [nm]
- $e_b$   black body flux [W/m$^2$]
- $f_v$   volume fraction of nanoparticles
- H    height of channel [m]
- $I_o$   initial intensity of radiation [W/m$^2$sr]
- $I$    intensity of radiation after travelling a distance '*s*' [W/m$^2$sr]
- K    coefficient of absorption or extinction
- $k$    conductivity [W m$^{-1}$ K$^{-1}$]
- $m$    normalized refractive index
- $n$    refractive index
- Q    flux [W/m$^2$]
- $Q_e$   extinction efficiency
- $q$    heat flux [W/m$^2$]
- *Re*   Reynolds Number

| | |
|---|---|
| *S* | solar irradiance [W/m$^2$] |
| *s* | distance travelled by radiation [m] |
| *T* | temperature [K] |
| *t* | time [sec] |
| *t$_1$* | thickness of the top plate of the channel |
| *u* | velocity in x-direction [m/sec] |
| *y* | position along channel depth [m] |

Greek Symbols:

| | |
|---|---|
| *α* | absorptivity |
| *β* | size parameter |
| *ε* | emissivity |
| *κ* | optical constant |
| *λ* | wavelength |
| *ϕ* | cosine of direction in which radiation is travelling |
| *ρ* | reflectivity |
| *θ* | optical depth |
| *τ* | transmissivity |
| *μ* | dynamic viscosity [kg m$^{-1}$ s$^{-1}$] |
| *ψ* | angle of incidence |

Subscript:

| | |
|---|---|
| *a* | absorption |
| *amb* | ambient |
| *avg* | average |
| *b* | black body |
| *e* | extinction |
| *eff* | effective |
| *f* | basefluid |
| *in* | inlet temperature |
| *loss* | loss |
| *nf* | nanofluid |
| *np* | nanoparticle |
| *sw* | solar weighted |
| *vac* | vacuum |
| *y* | position along channel depth |
| *1* | top plate of the channel |
| *2* | casing |
| *3* | bottom plate of the channel |
| *λ* | spectral |
| ⊥ | perpendicular |
| ∥ | parallel |

Superscript:

| | |
|---|---|
| *j* | number of reflections |
| + | direction of propagation from top to bottom |

- direction of propagation from bottom to top

Abbreviations and acronyms:
SAR         Surface absorption-based receiver
VAR         Volumetric absorption-based receiver
BS-HM       Black surface-heat mirror receiver design
G-A         Glass-atmosphere receiver design
G-G         Glass-glass receiver design
G-HM        Glass-heat mirror receiver design
SSS-ATM     Solar selective surface-atmosphere receiver design
SSS-G       Solar selective surface-glass receiver design